\documentclass[fleqn,usenatbib]{mnras}

\usepackage{newtxtext,newtxmath}

\usepackage[T1]{fontenc}

\DeclareRobustCommand{\VAN}[3]{#2}
\let\VANthebibliography\thebibliography
\def\thebibliography{\DeclareRobustCommand{\VAN}[3]{##3}\VANthebibliography}

\usepackage{graphicx}	
\usepackage{amsmath}	
\usepackage{lipsum}
\usepackage{booktabs}
\usepackage{array}
\usepackage{enumitem}
\usepackage{stfloats}

\graphicspath{{./figures/}}

\newif\ifedits
\newif\ifsecondedit
\newif\iffinaledits
\editsfalse
\secondeditfalse
\finaleditsfalse
\ifedits
    \newcommand{\edit}[1]{{\color{red} #1}}
    
\else
    \newcommand{\edit}[1]{#1}
    
\fi
\ifsecondedit
    \newcommand{\secondedit}[1]{{\color{red} #1}}
    
\else
    \newcommand{\secondedit}[1]{{\color{black} #1}}
    
\fi
\iffinaledits
    \newcommand{\finaledit}[1]{{\color{red} #1}}
    
\else
    \newcommand{\finaledit}[1]{{\color{black} #1}}
    
\fi

\newcommand\JWST{{\it JWST}}

\defcitealias{Mannucci_2010}{M10}
\defcitealias{Garcia_2024b}{Paper~I}
\defcitealias{Curti_2020}{C20}

\newcommand{\betterorder}[1]{}

\newcolumntype{x}[1]{>{\arraybackslash\hspace{0pt}}p{#1}}

\title[Does the FMR evolve with redshift? II]{Does the Fundamental Metallicity Relation Evolve with Redshift? II: The Evolution in Normalisation of the Mass-Metallicity Relation}

\author[Garcia et al.]{\ignorespaces
    Alex M. Garcia$^{1}$\thanks{E-mail: alexgarcia@virginia.edu},
    Paul Torrey$^{1}$, 
    Sara L. Ellison$^{2}$,
    Kathryn Grasha$^{3,4,5}$\thanks{ARC DECRA Fellow},
    Qian-Hui Chen$^{3,4}$,\newauthor
    Z.S. Hemler$^{6}$,
    Dhruv T. Zimmerman$^{7}$,
    Ruby J. Wright$^{8}$,
    Henry R.M. Zovaro$^{3,4}$,
    Erica J. Nelson$^{9}$,\newauthor
    Ryan L. Sanders$^{10}$,
    Lisa J. Kewley$^{11}$,
    Lars Hernquist$^{11}$
    \\~\\
    $^{1}$Department of Astronomy, University of Virginia, Charlottesville, VA 22904, USA\\
    $^{2}$Department of Physics \& Astronomy, University of Victoria, Finnerty Road, Victoria, British Columbia, V8P 1A1, Canada\\
    $^{3}$Research School of Astronomy \& Astrophysics, Australian National University, Canberra, Australia, 2611 \\
    $^{4}$ARC Centre of Excellence for All Sky Astrophysics in 3 Dimensions (ASTRO 3D), Australia\\
    $^{5}$Visiting Fellow, Harvard-Smithsonian Center for Astrophysics, 60 Garden Street, Cambridge, MA 02138, USA\\
    $^{6}$Department of Astrophysical Sciences, Princeton University, Peyton Hall, Princeton, NJ, 08544, USA \\
    $^{7}$Department of Astronomy, University of Florida, 211 Bryant Space Sciences Center, Gainesville, FL 32611 USA\\
    $^{8}$Department of Physics, University of Helsinki, Gustaf H{\"a}llstr{\"o}min katu 2, FI-00014 Helsinki, Finland\\
    $^{9}$Department for Astrophysical and Planetary Science, University of Colorado, Boulder, CO 80309, USA\\
    $^{10}$Department of Physics and Astronomy, University of Kentucky, 505 Rose Street, Lexington, KY 40506, USA\\
    $^{11}$Institute for Theory and Computation, Harvard-Smithsonian Center for Astrophysics, Cambridge, MA 02138, USA
}

\date{Accepted XXX. Received YYY; in original form ZZZ}

\pubyear{\the\year}

\begin{document}
\label{firstpage}
\pagerange{\pageref{firstpage}--\pageref{lastpage}}
\maketitle

\begin{abstract}
The metal content of galaxies is a direct probe of the baryon cycle.
A hallmark example is the relationship between a galaxy's stellar mass, star formation rate (SFR), and gas-phase metallicity: the Fundamental Metallicity Relation (FMR).
While low-redshift ($z\lesssim4$) observational studies suggest that the FMR is redshift-invariant, recent \secondedit{high-$z$} \JWST{} data indicate deviations \finaledit{from} \secondedit{the FMR established at low-$z$}.
In this study, we utilize the FMR to predict the evolution of the normalisation of the mass-metallicity relation (MZR) using the cosmological simulations Illustris, IllustrisTNG, EAGLE, and SIMBA.
Our findings demonstrate that a $z = 0$ calibrated FMR struggles to predict the evolution in the MZR of each simulation.
To quantify the divergence of the predictions, we introduce the concepts of a ``static'' FMR, where the role of the SFR in setting the normalization of the MZR does not change with redshift, and a ``dynamic'' FMR, where the role of SFR evolves over time.
We find static FMRs in SIMBA and dynamic FMRs in \secondedit{Illustris,} IllustrisTNG and EAGLE.
We suggest that the differences between these models likely points to the subtle differences in the implementation of the baryon cycle.
Moreover, we echo recent \JWST{} results at $z > 4$ by finding significant offsets from the FMR in IllustrisTNG and EAGLE, suggesting that the observed FMR may \secondedit{have a similar} dynamic \secondedit{trend as these simulations}.
Overall, our findings imply that the current FMR framework neglects important \edit{time} variations \edit{of these simulations'} baryon cycles.
\end{abstract}




\section{Introduction}
\label{sec:intro}

The baryon cycle encompasses the complete set of interactions involving baryonic matter within galaxies and their surrounding environments \citep[e.g.,][]{Oppenheimer_Dave_2006,Somerville_Dave_2015,Angles-Alcazar_2017,Tumlinson_2017,Wright_2024}.
Among the interactions within the baryon cycle are pristine gas accretion from the intergalactic medium (IGM, \citeauthor{Keres_2005} \citeyear{Keres_2005}; \citeauthor{Dekel_2006} \citeyear{Dekel_2006}), gas recycling from the circumgalactic medium (CGM, \citeauthor{Lacey_1985} \citeyear{Lacey_1985}; \citeauthor{Koeppen_1994} \citeyear{Koeppen_1994}), cold gas in the interstellar medium (ISM) collapsing to form stars \citep[e.g.,][]{McKee_2007,Kennicutt_2012}, stars synthesising heavy elements and expelling them back into the ISM via stellar feedback (through supernova explosions or asymptotic giant branch winds; \citeauthor{Friedli_1994} \citeyear{Friedli_1994}), and stellar feedback driving gas mixing/turbulence \citep[][]{Elmegreen_1999} as well as outflows \citep{Veilleux_2020}.
The net result of these processes (among others) working in tandem is a complex system that sustains galactic evolution.

Understanding the baryon cycle requires an observable probe of the underlying physics.
The metallicity of a galaxy offers one such reliable method of tracing galaxy evolution since metals have bright emission spectra (see, e.g., \citeauthor{Kewley_2019} \citeyear{Kewley_2019} and references therein) and act as tracers of the gaseous flows (\citeauthor{Lacey_Fall_1985} \citeyear{Lacey_Fall_1985}; \citeauthor{Schonrich_2009} \citeyear{Schonrich_2009}; see a recent review by \citeauthor{Tumlinson_2017} \citeyear{Tumlinson_2017}).

One example of the utility of metals as probes of the physics of galaxy evolution is found in the stellar mass-gas-phase metallicity relation (MZR).
The MZR describes an increasing relationship between the stellar mass of a galaxy and its gas-phase metal content \citep[see, e.g.,][]{Tremonti_2004,Lee_2006}.
The shape of the MZR is well-described by a power-law relation at low-to-intermediate masses ($M_* \lesssim 10^{10.5}$) with a flattening in the highest mass bins \citep[][]{Tremonti_2004,Zahid_2011,Berg_2012,Andrews_Martini_2013,Blanc_2019,Revalski_2024}.
\cite{Ellison_2008} first observed the existence of a secondary dependence within the scatter of the MZR -- galaxies with lower star formation rates (SFRs) tend to have higher gas-phase metallicity, and vice versa.
The secondary dependence on SFR can be understood in the context of the baryon cycle as the interplay of gas accretion and star formation \citep[e.g.,][etc]{Dave_2011,Dayal_2013,Lilly_2013,DeRossi_2015,Torrey_2018}.
Pristine gas accretes onto a galaxy from the IGM, increasing the gas fractions and lowering the metallicity.
The newly accreted gas is steadily consumed, indicated by increased star formation rates, producing more stars which, in turn, synthesise more metals.
The result of the gas consumption leaves the galaxy with lower gas fractions, lower SFRs, and higher metallicities.
Thus, at a fixed stellar mass, the gas-phase metallicity is anti-correlated with the gas content \citep{Bothwell_2013,Bothwell_2016,Torrey_2018} and SFR \citep{Ellison_2008,Lara_Lopez_2010,Mannucci_2010,Mannucci_2011} of galaxies (though at the highest stellar masses the relationship seemingly inverts; see, e.g., \citeauthor{Yates_2012} \citeyear{Yates_2012}).

The MZR's secondary dependence on star formation has been seen to persist at higher redshift \citep[e.g.,][]{Mannucci_2010, Belli_2013, Salim_2015, Sanders_2018, Sanders_2021}.
\cite{Mannucci_2010} claim that a single plane can describe the secondary dependence with SFR out to $z\!\sim\!2.5$.
The relationship that characterises this plane is a two-dimensional projection of the three-dimensional stellar mass, gas-phase metallicity, and SFR ($M_*-Z_{\rm gas}-{\rm SFR}$) relationship via
\begin{equation}
    \label{eqn:muAlpha}
    \mu_\alpha = \log M_* - \alpha \log {\rm SFR}~,
\end{equation}
a linear combination of the stellar mass and SFR.
In Equation~\ref{eqn:muAlpha}, $\alpha$ is a free parameter that is tuned to minimise the scatter in the 2D projection space.
\cite{Mannucci_2010} show that $\alpha=0.32$ describes galaxy populations out to $z\!\sim\!2.5$.
As such, many refer to this seemingly redshift-invariant (at least at $z\lesssim2.5$) relation as the ``Fundamental Metallicity Relation'' (FMR).

The FMR is customarily prescribed at $z=0$ and then extrapolated to higher redshift galaxy populations \citep[see][etc]{Mannucci_2010,Troncoso_2014,Wuyts_2014,Onodera_2016,Cresci_2019,Sanders_2021,Nakajima_2023}.
At $z\sim3$, galaxies at first appeared to be systematically metal poor compared to low redshift-calibrated FMR predictions \citep[][]{Mannucci_2010,Troncoso_2014,Onodera_2016}.
More recently, \cite{Sanders_2021} showed that the offsets at $z\!\sim\!3$ were due to both selection effects and choice of metallicity calibration.
Yet, recent \JWST{} observations have raised the issue again in finding offsets from the FMR at $z\gtrsim5$ \citep[][]{Heintz_2022,Langeroodi_2023,Nakajima_2023,Castellano_2024,Curti_2023}.
It is presently unclear whether these $z>5$ offsets are real deviations from the FMR, systematic uncertainties in \secondedit{the} derived metallicities, or detection limits from observations.

Furthermore, as we show in \cite{Garcia_2024a,Garcia_2024b}, it is not clear whether a single $\alpha$ value actually can characterise the scatter about the MZR at different redshifts in simulations in either the gas-phase or stellar metallicities.
In \citeauthor{Garcia_2024b} (\citeyear{Garcia_2024b}; henceforth also \citetalias{Garcia_2024b}), we define the idea of a ``strong'' FMR -- one in which a single $\alpha$ value can describe the scatter at all redshifts -- and a ``weak FMR'' -- one where a single $\alpha$ cannot describe the scatter at all redshifts.
We compared three widely-used cosmological simulations (Illustris, TNG, and EAGLE) and found that they exhibit weak FMRs \citep{Garcia_2024b}.
The FMR of observations appears to be ``strong'' at low redshift \citep[$z\lesssim3.5$;][]{Mannucci_2010,Sanders_2018,Sanders_2021}.
Yet, it should be noted that $\alpha$ values have been derived that seem to deviate with the \cite{Mannucci_2010} $\alpha$ value of 0.32 (see \citeauthor{Yates_2012} \citeyear{Yates_2012}, \citeauthor{Sanders_2017} \citeyear{Sanders_2017}, \citeauthor{Curti_2020} \citeyear{Curti_2020}, \citeauthor{Li_2023} \citeyear{Li_2023}, \citeauthor{Pistis_2024} \citeyear{Pistis_2024}, \citeauthor{Stephenson_2024} \citeyear{Stephenson_2024}).
This apparent disagreement is in part due to the derived $\alpha$ value depending on the chosen metallicity diagnostic \citep{Andrews_Martini_2013}.
Regardless, \cite{Sanders_2021} show that the FMR is redshift-invariant at least out to $z\sim3.5$ using self-consistent methodology.
This redshift-invariance is tied more closely to the offsets from the $z=0$ FMR \secondedit{predictions} rather than with the scatter about the MZR, however.

\edit{The FMR also makes predictions for the scatter about the MZR, however.}
It has been observed that the overall normalisation of the MZR decreases with increasing redshift \citep[e.g.,][]{Savaglio_2005,Maiolino_2008,Langeroodi_2022}.
Since galaxies at higher redshifts have higher SFRs \citep[see, e.g.,][]{Daddi_2007,Noeske_2007,Chen_2009,Speagle_2014,Koprowski_2024}, it is possible that the anti-correlation between SFRs and metallicities extends across redshift bins and drives the evolution of the normalisation in the MZR.
Yet, the implications of FMR \edit{has for the normalisation of the MZR} are not well understood or studied.

The importance of decoupling the two separate FMR predictions for scatter and normalisation cannot be overstated.
If the high-redshift offsets seen with recent \JWST{} observations are to believed, it is unclear whether they indicate the scatter about the MZR changing, or whether they are to do with the normalisation of the MZR.
We aim to provide a concrete framework for answering this question by separating out each prediction independently.
The primary aim of this paper is therefore to close the loop on the investigation started in \citetalias{Garcia_2024b}.
\edit{Whereas \citetalias{Garcia_2024b} focuses solely on the scatter,} here we focus solely on the FMR's predictions for the evolution of the normalisation of the MZR.
We ask: can a $z=0$ calibrated FMR can predict metallicities across cosmic time \edit{in cosmological simulations}?

The structure of this paper is as follows:
in Section~\ref{sec:methods} we overview the cosmological simulations Illustris, IllustrisTNG, EAGLE, and SIMBA, outline our galaxy selection criteria, and set-up an analytic framework for testing the MZR normalisation evolution.
In Section~\ref{sec:results}, we gather all of the necessary ingredients for predicting MZR evolution and then compare against each simulation's true MZR.
In Section~\ref{sec:discussion}, we present an alternative way of calibrating the FMR using high-redshift data and discuss the challenges of reproducing our methodology in observations.
Section~\ref{subsec:taking_stock} offers our view on the title question ``Does the FMR evolve with redshift?'' in the context of these findings as well as those of \citetalias{Garcia_2024b} as well as investigate the implications for recent observations.
Finally, Section~\ref{sec:conclusions} states our conclusions.

\section{Methods}
\label{sec:methods}

We use data products from the Illustris, IllustrisTNG, EAGLE, and SIMBA cosmological simulations.
The advantage of using multiple simulation models is that any common predictions between them should be independent of the detailed physical implementations.
Each of the simulations analysed here have fairly distinct implementations, yet there are a number of commonalities between the different models.
For example, each of the simulations analysed here have comparable box sizes as well as a sub-grid ISM pressurisation prescription.
We refer the reader to \citeauthor{Wright_2024} (\citeyear{Wright_2024}, their Table 1) for a succinct comparison of the EAGLE, TNG, and SIMBA physical models.
We furthermore include any notable difference between the implementations of the Illustris and TNG models not included in that reference (see second paragraph of Section~\ref{subsec:TNG}).

\subsection{Simulation Details}
\label{subsec:simulations}

\subsubsection{Illustris}
\label{subsec:Illustris}

Illustris \citep[][]{Vogelsberger_2013,Vogelsberger_2014a,Vogelsberger_2014b,Genel_2014,Torrey_2014} is a suite of cosmological hydrodynamic simulations run on the moving-mesh code {\sc arepo} \citep{Springel_2010}.
The Illustris physical model accounts for a range of (astro)physical processes, including gravity, star formation, stellar evolution, chemical enrichment, radiative cooling and heating of the ISM, stellar feedback, black hole growth, and active galactic nuclei (AGN) feedback.

The star-forming ISM is treated with an effective equation of state owing to the finite resolution of the Illustris simulations \citep{Vogelsberger_2013}.
This effective equation of state, \cite{Springel_Hernquist_2003}, allows star particles to form stochastically in the dense ($n_{\rm H} > 0.13~{\rm cm}^{-3}$) ISM.
A \cite{Chabrier_2003} initial mass function is assumed and the stellar evolutionary tracks follow from \citeauthor{Portinari_1998} (\citeyear{Portinari_1998}), which depend on the mass and metallicity of the star.
The Illustris framework models the return of mass and metals to the ISM through asymptotic giant branch (AGB) winds and Type II supernovae (SNe).
Illustris explicitly tracks nine species of chemical elements: H, He, C, N, O, Ne, Mg, Si, and Fe.
The yields of these elements are based on \cite{Karakas_2010} for AGB stars, \cite{Portinari_1998} for core-collapse SNe, and \cite{Thielemann_2003} for Type Ia SNe.
Relevant for this work, Illustris allows the diffusion of metals between gas elements using a gradient extrapolation scheme \citep{Vogelsberger_2014a}

The original Illustris suite is comprised of three $(75h^{-1}~{\rm Mpc})^3$ boxes with varying resolutions.
For this analysis, we use the data products from the highest resolution run, Illustris-1 (hereafter synonymous with Illustris).
The high resolution run has $2\times1820^3$ particles and an initial baryon mass resolution of $1.26\times10^6 M_\odot$.

\subsubsection{IllustrisTNG}
\label{subsec:TNG}

The Next Generation \citep[TNG;][]{Marinacci_2018,Naiman_2018,Nelson_2018,Pillepich_2018b,Springel_2018,Pillepich_2019, Nelson_2019a, Nelson_2019b} of the Illustris suite provides an update to the original model.
The Illustris and TNG models are therefore quite similar in spirit.
The unresolved star-forming ISM is treated with the same \cite{Springel_Hernquist_2003} effective equation of state (at the same threshold density).
Star particles assume the same \cite{Chabrier_2003} IMF.
The stellar lifetime models of TNG are also adopted from \cite{Portinari_1998}.
TNG explicitly tracks the same nine chemical species (with a tenth, ``other metals'', added as a proxy for metals not explicitly tracked) and metals are advected between gas elements using a similar gradient extrapolation scheme \citep{Pillepich_2018a}.

There are notable difference between Illustris and TNG, however \citep[see][for full comparison of the models]{Weinberger_2017,Pillepich_2018a}.
Changes between Illustris and TNG that are relevant to the \cite{Wright_2024} reference table are:
(i) a slightly higher initial baryon mass resolution in Illustris\ignorespaces
\footnote{
The difference in mass resolution is entirely dependent on the changed value of $h$ (the reduced Hubble Parameter) between the different simulations.
Illustris uses WMAP \citep{Hinshaw_2013} cosmology with $h=0.704$ while TNG uses Planck \citep{Planck_2016} cosmology with $h=0.677$
}, 
(ii) the inclusion of magnetic fields in TNG, 
(iii) the TNG model introducing redshift-scaling winds making feedback more efficient at high-redshift, and 
(iv) low-accretion rate AGN feedback in the Illustris model being thermal `bubbles' versus black hole driven kinetic winds in the TNG model.
Another relevant change is the updated yield tables: TNG gets its elemental yields from \cite{Nomoto_1997} for Type Ia SNe, \cite{Portinari_1998} and \cite{Kobayashi_2006} for Type II SNe and as \cite{Karakas_2010}, \cite{Doherty_2014}, and \cite{Fishlock_2014} for AGB stars.
Moreover, the minimum mass for core-collapse SNe is increased from $6M_\odot$ in Illustris to $8M_\odot$ in TNG.
As one example of the impact of one these changes between Illustris and TNG, we show that the redshift-dependent wind implementation in the TNG model plays a significant role in setting the importance of SFRs in determining both stellar \citep{Garcia_2024a} and gas-phase metallicities \citep{Garcia_2024b} compared to the original Illustris simulation.

The suite of TNG simulations ranges from boxes of size $(35h^{-1}~{\rm Mpc})^3$ to $(205h^{-1}~{\rm Mpc})^3$.
As a fair comparison with the selected Illustris run, we use data products from the highest-resolution $(75h^{-1}~{\rm Mpc})^3$ box, TNG100-1 (hereafter simply TNG), which has $2\times1820^3$ particles and an initial baryon mass resolution of $1.4\times10^6~M_\odot$.

\subsubsection{EAGLE}
\label{subsec:EAGLE}

The ``Evolution and Assembly of GaLaxies and their Environment'' \citep[EAGLE;][]{Crain_2015, Schaye_2015, McAlpine_2016} simulations are built on a heavily modified version of {\sc gadget-3} (\citeauthor{Springel_2005} \citeyear{Springel_2005}) smoothed particle dynamics (SPH) code called {\sc anarchy} (see \citeauthor{Schaye_2015} \citeyear{Schaye_2015}, their Appendix A).
Much like Illustris and TNG, EAGLE models the physics of gravity, stellar formation, evolution and feedback, radiative cooling and heating of the ISM, as well as black hole growth and AGN feedback.

EAGLE also has finite resolution and treats the unresolved star-forming ISM with an effective equation of state \citep{Schaye_DallaVechhia_2008}.
The \cite{Schaye_DallaVechhia_2008} and \cite{Springel_Hernquist_2003} equations of state are qualitatively similar, yet they have differences in their detailed implementations.
For example, the \cite{Schaye_DallaVechhia_2008} model has additional metallicity- and temperature-dependent criteria for the density threshold of star formation (see \citeauthor{Schaye_2004} \citeyear{Schaye_2004} and \citeauthor{Schaye_2015} \citeyear{Schaye_2015}).
Similarly, while both use a \cite{Chabrier_2003} IMF, the \cite{Schaye_DallaVechhia_2008} models its stellar evolution based on the \cite{Wiersma_2009b} model.
EAGLE explicitly tracks the same nine chemical species as Illustris in addition to Sulfur and Calcium.
The yields of these elements come from \cite{Portinari_1998} for Type II SNe, \cite{Marigo_2001} for AGB stars, and \cite{Wiersma_2009b} and \cite{Thielemann_2003} for Type Ia SNe.
Notably, since EAGLE implements SPH (which is Lagrangian), metals in EAGLE are not advected to adjacent gas elements as they are in Illustris and TNG.

The suite of EAGLE simulations ranges from boxes of size $(12.5~{\rm Mpc})^3$ to $(100~{\rm Mpc})^3$. 
As an even-handed comparison with the selected Illustris and TNG runs, we use data products from the $(67.8h^{-1}~{\rm Mpc})^3$ (or $[100~{\rm Mpc}]^3$) box also known as {\sc RefL0100N1504} (hereafter simply EAGLE).
This run of EAGLE has $2\times1504^3$ particles and initial baryon mass resolution of $1.81\times10^6 M_\odot$.

\subsubsection{SIMBA}
\label{subsec:SIMBA}

The SIMBA \citep{Dave_2019} simulations are the successor to the MUFASA simulations \citep{Dave_2016}.
SIMBA is built using the meshless finite mass (MFM) version of {\sc gizmo} \citep{Hopkins_2015,Hopkins_2017} which itself is based on the {\sc gadget}-3 \citep{Springel_2005} code.
SIMBA models many of the same astrophysics as the other simulations; however, a notable addition is the inclusion of dust production, growth and destruction \citep[see, e.g.,][]{Li_2019}.
Therefore, some of the metals in SIMBA are locked in the dust (following the prescription of \citeauthor{Dwek_1998} \citeyear{Dwek_1998}).
We note that the metals converted to dust are not considered in the gas-phase metallicities reported in this analysis.

Similar to all the other simulations analysed in this work, SIMBA employs a sub-resolution prescription for the unresolved ISM.
SIMBA uses a molecular hydrogen-based star formation prescription based on \cite{Krumholz_2011}.
The H$_2$ fraction threshold for star formation has criteria set by both the metallicity and local column density of gas \citep{Krumholz_2011}.
Gas meeting this H$_2$ criterion forms new star particles, which eventually return their mass and metals to the ISM in a manner consistent with the mass loss of a stellar population characterized by a \cite{Chabrier_2003} initial mass function.
SIMBA tracks the same 11 chemical species as EAGLE.
The elemental yields for SIMBA come from \cite{Nomoto_2006} for Type II SNe, \cite{Iwamoto_1999} for Type Ia SNe, and \cite{Oppenheimer_Dave_2006} for AGB stars.
Similar to the SPH method implemented in EAGLE, the MFM (Lagrangian) nature of {\sc gizmo} means that metals are not diffused to neighbouring gas elements in SIMBA.
Important for this the discussion later in this work, SIMBA implements star formation-driven outflows as a decoupled two-phase wind.
The mass loading factors of these winds ($\eta$) are determined based on work in the Feedback in Realistic Environments (FIRE) zoom-in simulations tracking individual particles \citep{Angles_Alcazar_2017b}.
A manual suppression of $\eta$ is applied at high redshifts ($z>3$) to allow for early galaxy growth in poorly resolved situations.

The SIMBA suite has simulations with box sizes ranging from $(25h^{-1} ~{\rm Mpc})^3$ to $(100h^{-1}~{\rm Mpc})^3$.
We use data from the largest box for the most fair comparison to the selected volumes of Illustris, TNG, and EAGLE.
The inital baryon mass resolution of the highest resolution $(100h^{-1}~{\rm Mpc})^3$ SIMBA box is $1.82\times10^7M_\odot$, \secondedit{corresponding to} $1024^3$ particles.
We note that the baryon mass resolution is an order of magnitude coarser than the other three simulations analysed here.

\subsection{Galaxy Selection Criteria}
\label{subsec:galaxy_selection}

Substructure in \secondedit{each simulation} is found using {\sc subfind} \citep{Springel_2001,Dolag_2009}.
{\sc subfind} 
identif\secondedit{ies} gravitationally-bound collections of particles using a friends-of-friends (FoF; \citeauthor{Davis_1985} \citeyear{Davis_1985}) algorithm.
All of the physical parameters in this work are from the {\sc subfind} 
galaxy catalogs \secondedit{(with the notable exception of the direct tracking of the oxygen and hydrogen species in Appendix~\ref{appendix:oxygen_abundances})}.

We limit our analysis to ``well-resolved'' central galaxies in the simulations.
We define well-resolved as having $\sim\!10^2$ star particles (stellar mass of $>10^8 M_\odot$ in Illustris, TNG, and EAGLE, $>10^9 M_\odot$ in SIMBA) and $\sim\!5\times10^2$ gas particles (gas mass of $>10^{8.5} M_\odot$ in Illustris, TNG and EAGLE, $>10^{9.5}$ in SIMBA).

We furthermore only select star-forming galaxies from our simulations in order to make as fair a comparison as possible to observations.
Metallicity is derived from emission lines coming from star-forming regions of galaxies in observations \citep[][and references therein]{Kewley_Ellison_2008,Kewley_2019}.
Derived metallicity values are therefore typically limited to star-forming regions within star-forming galaxies.
Our star-forming galaxy cut is based on a galaxy's position relative to the specific star formation main sequence (sSFMS).
We define the sSFMS as a median specific star formation rate-stellar mass relationship in mass bins of 0.05 dex for galaxies with stellar masses $<10^{10.2} M_\odot$.
The sSFMS is a linear least-squares fit to this median relation, with extrapolation for galaxies above stellar mass of $10^{10.2}$.
Any galaxy that has a sSFR less than 0.5 dex below the sSFMS is omitted from this analysis\ignorespaces
\footnote{
We note that our key results are not sensitive to the choice in sSFMS, however (see Appendix~\ref{appendix:sSFMS}).
}.

\subsubsection{Derivation of Metallicity}
\label{subsubsec:derived_metallicities}

In observations, gas-phase metallicities are typically reported in units of
\begin{equation}
    \label{eqn:logO/H}
    \log ({\rm O/H})+12~.
\end{equation}
To convert to these observer units, we assume that the metal mass fraction of oxygen ($f_{\rm O}$) and total mass fraction of hydrogen ($X$) are constants.
We adopt values of \secondedit{$f_{\rm O} = 50\%$} and $X=76\%$ for this analysis.
In more detail, $f_{\rm O}$ depends on the sSFR of the galaxy: galaxies with high sSFR should be dominated by core-collapse enrichment increasing $f_{\rm O}$, whereas low sSFR galaxies should have more contributions from AGB and Type Ia SNe diluting $f_{\rm O}$.
Naturally, $f_{\rm O}$, and the evolution thereof, is sensitive to detailed yield tables from each of the simulations (see references for each simulation's yields in Section~\ref{subsec:simulations}).
$f_{\rm O}$ is consequently not constant in the simulations when tracked more carefully \secondedit{(nor, indeed, is $X_{\rm H}$; see Appendix~\ref{appendix:oxygen_abundances}, Figure~\ref{fig:oxygen_abundances})}.
However, we make the simplifying assumption of a constant $f_{\rm O}$ as the FMR is designed to trace the {\it metallicity} of galaxies, not the oxygen abundances specifically.
By tracking the oxygen abundances directly, we would be probing the differences in the elemental yields -- which are not uniform between the different simulation models.
\secondedit{We aim to mitigate the impact of the different yield tables by using (and scaling) the overall metallicity.
The overall metallicity is not entirely independent of the yield tables, but should be less sensitive than an individual species.
We \finaledit{explore} the decision to scale the overall metallicity further in Appendix~\ref{appendix:oxygen_abundances}.
Here we mention that the core results are not significantly impacted if we instead directly traced Oxygen and Hydrogen abundances from the simulations.
}
\secondedit{We therefore} caution that, while we report metallicity in units of Equation~\ref{eqn:logO/H}, we are not reporting directly tracked oxygen abundances from the simulation.

Finally, we limit all gas-phase metallicities derived from the simulations to that of \secondedit{a mass-weighted metallicity of all} star-forming gas (defined for Illustris/TNG in Section~\ref{subsec:Illustris}, EAGLE in Section~\ref{subsec:EAGLE}, and SIMBA in Section~\ref{subsec:SIMBA}) \secondedit{bound to the system}.
\secondedit{
We note that using either the SFR-weighted metallicity or mass-weighted metallicity of {\it all} gas instead has only a modest effect on the overall normalisation of the MZR, thus our choice of mass-weighted metallicity within star forming gas \finaledit{does} not significantly impact our conclusions.
}

\subsection{Combining the SFMS and FMR to predict the MZR}
\label{subsec:FMR_definitions}

\begin{figure*}
    \centering    
    \includegraphics[width=\linewidth,trim={4cm 0 0 0}]{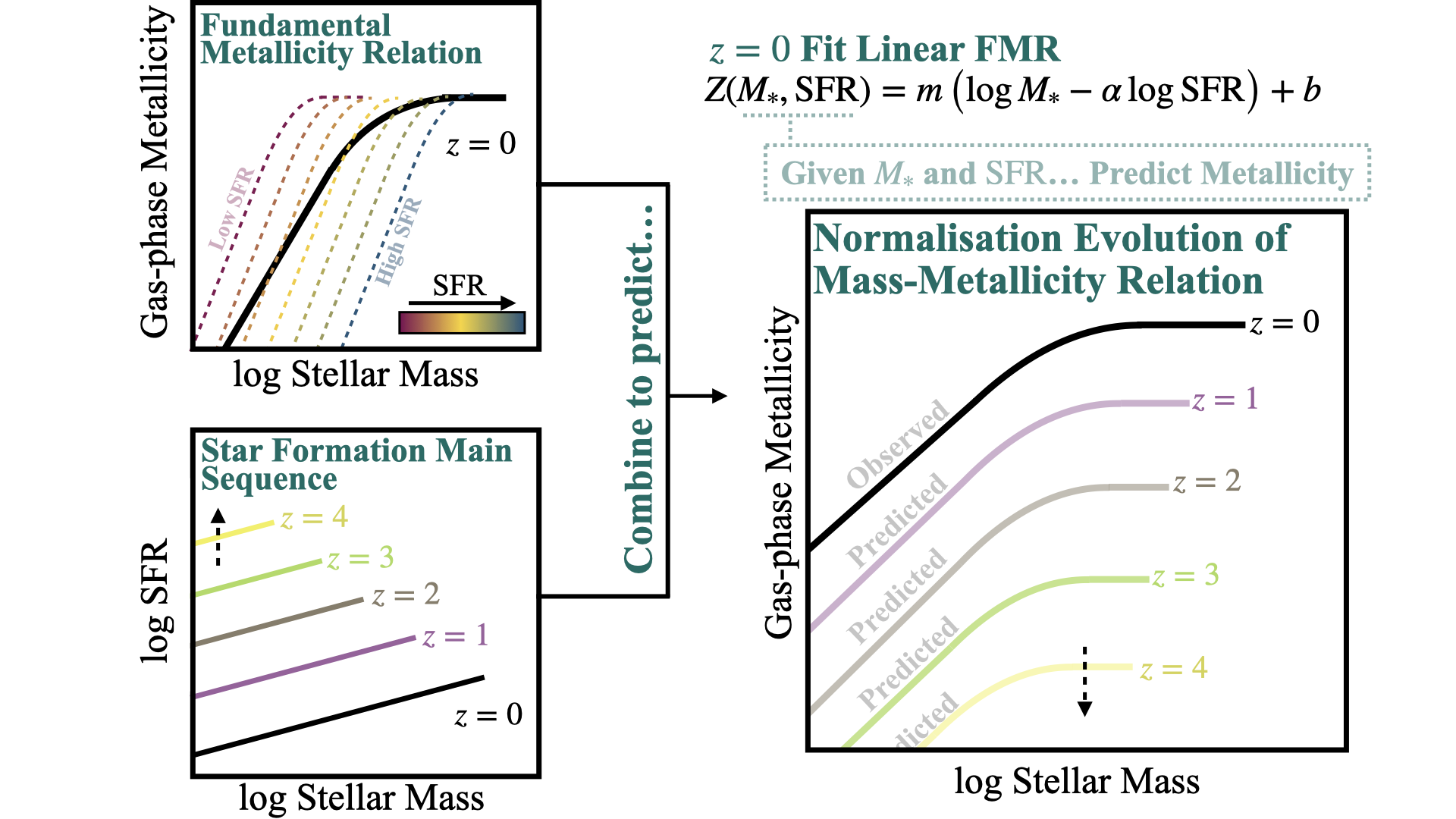}
    \caption{{\bf The Combination of the Fundamental Metallicity Relation and the Star Formation Main Sequence Can Predict Mass Metallicity Relations.} Schematic illustrating how the Fundamental Metallicity Relation (FMR; top left) and Star Formation Main Sequence (SFMS; bottom left) combine to predict the evolution in the normalisation of the stellar mass-gas-phase metallicity relation (MZR; right).
    The FMR functional form has mass and SFR as inputs and is typically fit using $z=0$ galaxy populations.
    Therefore, the addition of average masses and average SFRs at higher redshifts from the SFMS provide a prediction for the average MZR at higher redshifts.
    }
    \label{fig:cartoon}
\end{figure*}

The FMR is typically expressed as a relationship explaining the scatter about the MZR.
\edit{\ignorespaces
More specifically, this ``fundamental'' relation suggests galaxies that are metal poor compared to the MZR tend to have lower SFRs and metal rich galaxies tend to have higher SFRs.
The term fundamental is tied with the observed lack of variation out to $z\sim3.5$ \citep{Mannucci_2010,Sanders_2021}.
Recent \JWST{} observations seem to challenge this notion, however, suggesting that at $z\gtrsim5$ galaxies are more metal poor compared to FMR predictions.
Moreover, in \citetalias{Garcia_2024b}, we show that the detailed scatter about the MZR in simulations has some previously unaccounted for temporal deviations.
However, it is not clear that the deviations in the correlation between offsets from the MZR and SFR can necessarily explain the observed high-redshift offsets.
}
\edit{We should recall that the} FMR also makes predictions for the evolution of its normalisation.
\edit{\ignorespaces
The increased SFRs at higher redshifts (i.e., along the SFMS; \citeauthor{Noeske_2007} \citeyear{Noeske_2007}, \citeauthor{Daddi_2007} \citeyear{Daddi_2007}, etc.) may be responsible for the observed decrease in galactic metallicity at earlier times \citep{Savaglio_2005, Maiolino_2008, Zahid_2011, Langeroodi_2023}.
It is possible that changes in the behaviour of this aspect of the FMR are what drive the offsets at high redshift.
}
The language to describe the \edit{evolution in the normalisation} is less developed in the literature.
We therefore dedicate this section to developing the framework to explicitly relate the FMR to the redshift evolution of the MZR.

We first assume the common linear form of the FMR\ignorespaces
\footnote{
In principle this can be done for any functional form of the FMR.
We choose linear as it is most consistent with recent observational methodology \citep{Heintz_2022,Langeroodi_2023,Nakajima_2023,Castellano_2024, Curti_2023}.
See Appendix~\ref{appendix:different_FMRs} for a discussion of different regression methods.
}
such that
\begin{equation}
    \label{eqn:linear}
    Z = m \left(\log M_* - \alpha \log {\rm SFR}\right) + b~,
\end{equation}
where $m$ is the slope, $b$ is the $y$-intercept, and $\alpha$ is the free parameter from Equation~\ref{eqn:muAlpha} that produces the minimum scatter relation (see \citetalias{Garcia_2024b} for a full discussion of $\alpha$).
The median metallicity of a set of galaxies is set by the median SFR and median mass.
However, the median SFR itself is a function of stellar mass via the star formation main sequence (SFMS; \citeauthor{Noeske_2007} \citeyear{Noeske_2007}, \citeauthor{Daddi_2007} \citeyear{Daddi_2007}, etc).
Therefore the average metallicity at any redshift can be written such that
\begin{equation} 
    \label{eqn:MZR_full}
    \langle Z(M_*, z) \rangle =  m \left( \langle\log M_*\rangle - \alpha \log \langle {\rm SFR} (M_*, z) \rangle  \right) + b~,
\end{equation}
where $\langle {\rm SFR} (M_*, z) \rangle $ is the SFMS.
Equation~\ref{eqn:MZR_full} is an approximation to the MZR insofar as it describes the median metallicity as a function of stellar mass.

Therefore, given some observed evolution in the SFMS and a $z=0$ calibrated FMR, one can make predictions for what the normalisation of the MZR {\it should} be at any redshift (this idea illustrated in Figure~\ref{fig:cartoon}).
The primary aim of this paper is to investigate whether or not the preceding statement is true.
Does the FMR, calibrated at $z=0$, correctly predict the median MZR based on the evolution of the SFMS alone? 
We test the extent to which this logic holds \edit{in simulations} in the following sections.
\edit{\ignorespaces
We note that this analysis omits the scatter about the MZR.
This is to completely isolate the evolution of the normalisation of the MZR.
Moreover, this work focuses on the predictions of the FMR compared to the actual evolution of the MZR from the simulations, rather than a direct comparison to the $z=0$ relation.
For a detailed examination of the scatter about the MZRs of these simulations, and how it compares to the $z=0$ relations, see \citeauthor{Garcia_2024b} (\citeyear{Garcia_2024b}) for Illustris, TNG, and EAGLE, and Appendix~\ref{appendix:SIMBA_strong_weak} for SIMBA.}

\section{Results}
\label{sec:results}

\begin{figure*}
    \centering
    \includegraphics[width=\linewidth]{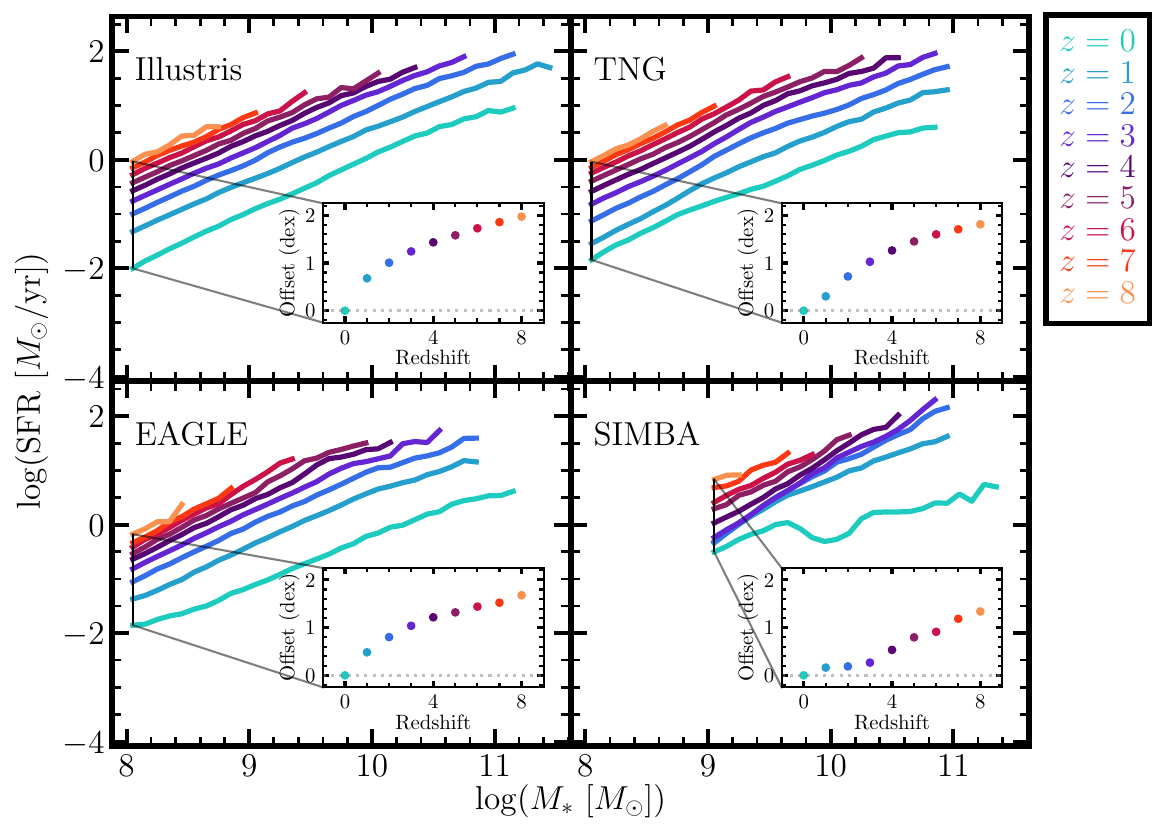}
    \caption{{\bf Evolution of the Star Formation Main Sequence (SFMS).}
    The median SFMS for resolved, star-forming galaxies in Illustris (top left), TNG (top right), EAGLE (bottom left), and SIMBA (bottom right).
    The SFMS is presented at $z=0-8$ in Illustris, TNG, and EAGLE.
    We note that we require at least 20 galaxies in each mass bin.
    SIMBA therefore only extends to $z=7$ and down to $\log(M_*~[M_\odot]) = 9.0$ since the SIMBA simulation analysed here has an order of magnitude coarser mass resolution than Illustris, TNG, and EAGLE.
    The insets in each panel show the offsets from the $z=0$ SFMS as a function of redshift in the lowest mass-bin of each simulation.
    }
    \label{fig:SFMS_Comp}
\end{figure*}

\begin{figure*}
    \centering
    \includegraphics[width=\linewidth]{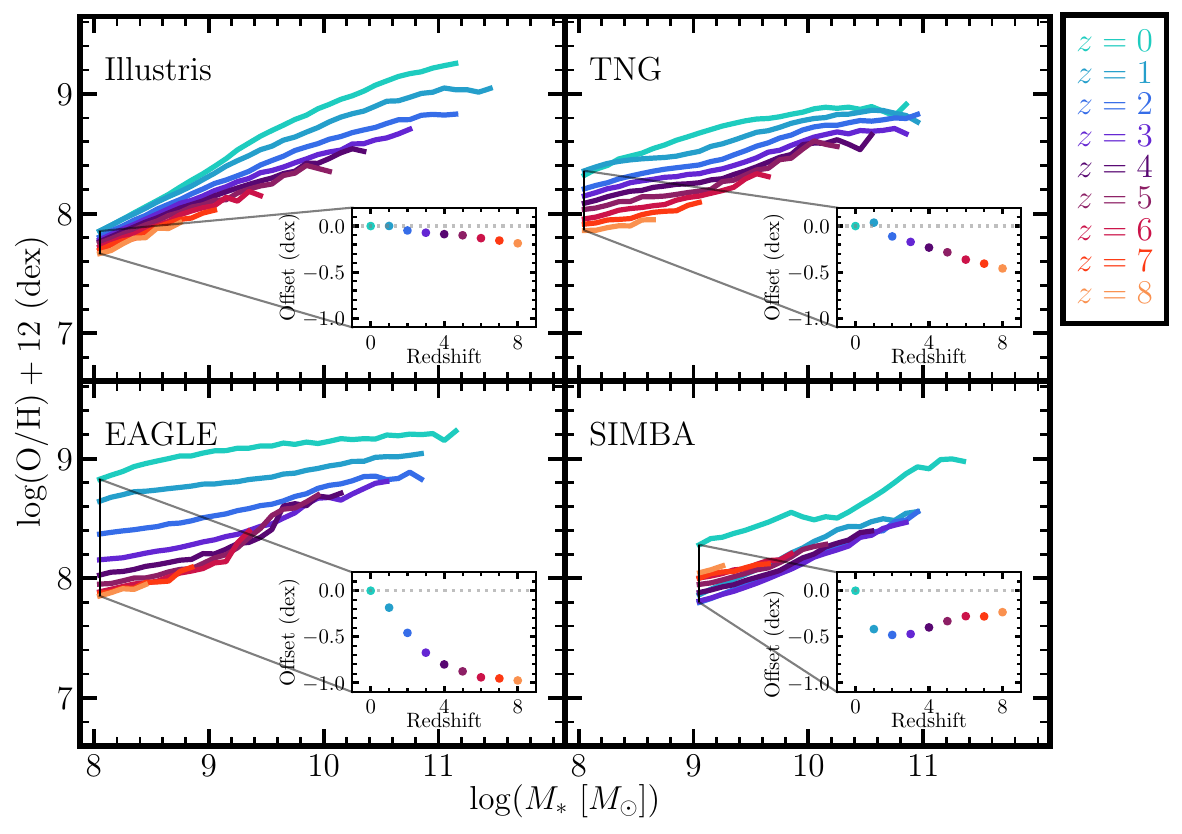}
    \caption{{\bf Evolution of the Mass Metallicity Relation (MZR).}
    The median MZR for resolved, star-forming galaxies in Illustris (top left), TNG (top right), EAGLE (bottom left), and SIMBA (bottom right).
    Similar to Figure~\ref{fig:SFMS_Comp}, MZRs are presented for $z=0-8$ for Illustris, TNG, and EAGLE.
    The MZR in SIMBA extends out to only $z=7$, and down to $\log(M_*~[M_\odot]) = 9.0$, owing to the lower resolution of the chosen SIMBA simulation.
    We again require at least 20 galaxies in each mass bin.
    The insets in each panel show the offsets from their $z=0$ MZR as a function of redshift in the lowest mass-bin of each simulation.
    }
    \label{fig:MZR_Comp}
\end{figure*}

\subsection{The Ingredients for Predicting MZR Evolution}
\label{subsec:normalisation}

The previous section details how predictions for the normalisation of the MZR can be made.
In this section, we gather each of the components of the prediction, set our point of comparison, and present our predictions.
Both the SFMS and MZR, as well as their respective evolution, have been presented for these simulations previously (for SFMS see \citeauthor{Furlong_2015} \citeyear{Furlong_2015}; \citeauthor{Sparre_2015} \citeyear{Sparre_2015}; \citeauthor{Donnari_2019} \citeyear{Donnari_2019}; \citeauthor{Akins_2022} \citeyear{Akins_2022}, for gas-phase MZR see \citeauthor{Torrey_2014} \citeyear{Torrey_2014}, \citeyear{Torrey_2019}; \citeauthor{DeRossi_2017} \citeyear{DeRossi_2017}; \citeauthor{Dave_2019} \citeyear{Dave_2019}).
We choose to present them again here as a direct and even-handed comparison of the predictions made by these models.

\subsubsection{The \texorpdfstring{$z=0$}{z=0} Calibrated FMR}
\label{subsubsec:z=0FMR}

Calibrating the FMR involves two steps: (i) determine the projection of least scatter via Equation~\ref{eqn:muAlpha} and (ii) fit a regression to the projected $\mu_\alpha\!-\!{\rm metallicity}$ relation.
We follow the same methodology as \citetalias{Garcia_2024b} (which is based on \citeauthor{Mannucci_2010} \citeyear{Mannucci_2010}) to compute the projection of least scatter.
We vary $\alpha$, the free parameter of Equation~\ref{eqn:muAlpha}, from 0 to 1 in steps of 0.01.
We fit each corresponding $\mu_\alpha\!-\!{\rm metallicity}$ relation with a linear regression.
Whichever $\alpha$ value produces the minimum scatter (smallest standard deviation of residuals) in the new $\mu_\alpha\!-\!{\rm metallicity}$ space is deemed $\alpha_{\rm min}$.
This $\alpha_{\rm min}$ parameter encodes the projection direction from 3D (mass-metallicity-SFR) space into 2D ($\mu_\alpha\!-\!{\rm metallicity}$) space such that the scatter in 2D is minimised.
We find that $\alpha_{\rm min}$ at $z=0$ is 0.23 in Illustris, 0.31 in TNG, 0.74 in EAGLE, and 0.33 in SIMBA (see \citetalias{Garcia_2024b} and Appendix~\ref{appendix:SIMBA_strong_weak}).

The regression to the FMR is the linear least-squares fit to the minimum scatter $\mu_{\alpha}\!-\!{\rm metallicity}$ distribution.
We choose to compare to the linear form in the main body of this work to 
(i) link more directly to the analytic framework derived in Section~\ref{subsec:FMR_definitions} as well as
(ii) to create an even-handed comparison to recent \JWST{} observational results \citep{Heintz_2022,Langeroodi_2023,Nakajima_2023,Castellano_2024,Curti_2023}.
Appendix~\ref{appendix:different_FMRs} offers a discussion of {different fitting methods}.
Our main conclusions are relatively insensitive to the choice of regression, however (see further discussion in Appendix~\ref{appendix:different_FMRs}).
The best fit regressions are
\secondedit{
\begin{equation}
\label{eqn:FMR_regressions}
\begin{alignedat}{3}
   &[\log {\rm O/H} + 12]_{\rm Illustris} &= 0.693(\mu_{0.23}) + 1.807~,&\\
   &[\log {\rm O/H} + 12]_{\rm TNG} &= 0.368(\mu_{0.31}) + 5.044~,&\\
   &[\log {\rm O/H} + 12]_{\rm EAGLE} &= 0.380(\mu_{0.74}) + 5.132~,&~{\rm and}\\
   &[\log {\rm O/H} + 12]_{\rm SIMBA} &= 0.364(\mu_{0.66}) + 4.901~.&\\
\end{alignedat}
\end{equation}
}
We find that the slopes, intercepts, and $\alpha_{\rm min}$ values of the FMR are quite different from simulation-to-simulation.
The discrepancy between the different models underscores that each simulation model presented here has a quantitatively different prescription for modeling baryonic physics.
While difficult to interpret directly from Equation~\ref{eqn:FMR_regressions} (and, indeed, any functional form of the FMR), we highlight some potential differences that may give rise to the different FMRs in Section~\ref{subsec:physics_of_FMR}.

We note that we take $\alpha_{\rm min}$ to be constant throughout this work (i.e., strong FMR).
Assuming a strong FMR makes the (incorrect, in the case of these simulations; see \citetalias{Garcia_2024b}) assumption that a single $\alpha_{\rm min}$ can describe galaxy populations across redshift in simulations.
We make this simplifying assumption as a test to the extent to which a $z=0$ calibrated FMR can reproduce the evolution in normalisation of the MZR.

\subsubsection{Evolution of the SFMS}
\label{subsubsec:SFMS}

We define the SFMS for each simulation as the median SFR in mass bins of width 0.1 dex.
Figure~\ref{fig:SFMS_Comp} shows this median SFMS for Illustris (top left), TNG (top right), EAGLE (bottom left), and SIMBA (bottom right).
We note that we require more than 20 galaxies in a mass bin and that our conclusions are largely unimpacted by reasonable changes in the definition of the median relation \secondedit{(e.g., adopted bin width or minimum number of galaxies)}.

We find that the SFMS is a power-law relationship at each redshift in each simulation (with the notable exception of $z=0$ in SIMBA, see discussion below).
The agreement between the simulations' SFMS is noteworthy.
For example, the normalisation and slope of the $z=0$ SFMS are quite similar in Illustris, TNG, and EAGLE: around $-2 ~{\rm dex}$ at $\log(M_*~[M_\odot]) = 8$ to $\sim0.5~ {\rm dex}$ at $\log(M_*~[M_\odot]) = 11$ giving a slope of $\sim0.9$.
In SIMBA, however, the slope is shallower at 0.4.
The power-law slopes of in the simulations are broadly consistent with observational slopes with SIMBA being on the shallower end (e.g., \citeauthor{Speagle_2014} \citeyear{Speagle_2014}; \citeauthor{Lee_2015} \citeyear{Lee_2015}; \citeauthor{Tomczak_2016} \citeyear{Tomczak_2016}).
There have been a number of studies that report a flattening of the SFMS at high masses \citep[e.g.,][]{Lee_2015,Leslie_2020,Popesso_2023,Koprowski_2024}.
Broadly speaking, the simulations analysed here do not have this same turnover at the highest masses.
\secondedit{The lack of turn-over in the SFMS at the highest masses is due to our star forming galaxy criteria (see Section~\ref{subsec:galaxy_selection}).
By taking less restrictive cuts (i.e., adding galaxies that are ``less'' star forming to the sample), there is indeed a turnover at the highest masses, though not at all redshifts (generally only at $z\lesssim2-3$).
}

We find that the overall normalisation of the SFMS increases in each simulation such that, in a fixed mass bin, the highest SFR galaxies are at the highest redshift (see insets on each panel of Figure~\ref{fig:SFMS_Comp}).
The behaviour of increasing normalisation at higher redshifts is qualitatively consistent with observations \citep[e.g.,][]{Daddi_2007,Noeske_2007,Chen_2009,Speagle_2014,Katsianis_2020}.
In more detail, there are some subtle differences in the redshift evolution in the SFMS between the four simulations (see inset panels of Figure~\ref{fig:SFMS_Comp}).
Illustris, TNG, and EAGLE all have evolution that is larger at low redshift and smaller at higher redshift. 
Despite this similar trend, Illustris has more evolution from $z=0$ to $z=1$ than TNG and EAGLE.
The evolution past $z=1$ is roughly linear in Illustris; however, this linear evolution is not apparent in TNG and EAGLE until $z\sim3$.
SIMBA, on the other hand, seems to have more evolution at higher redshift in the lowest mass bins.
In fact, in this lowest mass bin there is virtually no evolution in the SFMS from $z=1$ to $z=2$\ignorespaces
\footnote{
At higher masses, however, e.g., $\log (M_*~[M_\odot])\gtrsim10$, there does appear to be {\it some} redshift evolution between $z=1$ and $z=2$.
}.
Interestingly, SIMBA also seems to have roughly linear evolution at $z\gtrsim3$.
In summary, while there are indeed subtle differences between each simulation the SFMS (and its evolution) is broadly consistent between the four simulation models.

\subsubsection{Evolution of the MZR}
\label{subsubsec:MZR}

With the $z=0$ FMR and evolution of SFMS, we can now make predictions for the average evolution of the MZR.
However, it is worth first setting a reference for comparison with the predictions.
We define the median MZR in an analogous way to the median SFMS: a median metallicity of the star-forming gas in stellar mass bins of 0.1 dex.
We again require there to be at least 20 galaxies in each mass bin.
As with our definition of the median SFMS, we note that our key results are relatively insensitive to the specific definition of the MZR.

Figure~\ref{fig:MZR_Comp} shows the evolution of the MZR for Illustris (top left), TNG (top right), EAGLE (bottom left), and SIMBA (bottom right).
Whereas the broad agreement between the SFMS in each simulation was noteworthy in the last section, here the {\it dis}agreement between the MZRs (and their respective evolution) is remarkable.
In Illustris at $z=0$, we find a relatively steep MZR that is roughly linear with a turnover at $\log (M_*~[M_\odot])\sim10$.
Both TNG and EAGLE have much shallower MZRs at $z=0$ with the former having a turnover mass of about $\log (M_*~[M_\odot])\sim10$ and the latter not having any clear turnover mass.
The SIMBA $z=0$ MZR exhibits a similar steepness to that of Illustris, although it lacks a significant turnover.
Overall, there is significant disparity between the four different simulations at $z=0$.
The significantly different $z=0$ MZRs supports the point made that FMR regression parameters (e.g., Equation~\ref{eqn:FMR_regressions}) vary significantly from simulation-to-simulation.

Furthermore, each simulation makes profoundly different predictions for the evolution in the normalisation of the MZR (see inset panels of Figure~\ref{fig:MZR_Comp}).
There is minimal evolution in Illustris: $\lesssim0.2$ dex from $z=0$ to $z=8$ in the lowest mass bin.
The evolution in this lowest mass bin in practically linear with redshift.
TNG shows comparatively more evolution ($\sim$0.5 dex from $z=0$ to $z=8$ in the lowest mass bin) and is similarly linear in redshift space.
EAGLE shows a quite large amount of redshift evolution -- $\sim$1 dex from $z=0$ to $z=8$ in the smallest mass bin.
Interestingly, the evolution in EAGLE is non-linear with redshift with the majority occurring at $z\lesssim5$.
Finally, in SIMBA, the normalisation of the MZR in the lowest mass bin decreases at $z<4$ and then, surprisingly, proceeds to increase out to $z=7$ (at which point it is virtually consistent the $z=0$ metallicity).
This behaviour is consistent at higher mass bins, as well.
On either side of this normalisation ``turnaround'' the evolution is roughly linear.

One possible explanation for the the inversion of the MZR at high redshift in SIMBA is the suppression of winds at high-redshift.
As was mentioned in Section~\ref{subsec:SIMBA}, the mass loading factors of winds, $\eta$, are suppressed at high-$z$ in the SIMBA model to allow for low-mass galaxy growth in these early epochs (\citeauthor{Dave_2019} \citeyear{Dave_2019}; which was noted to be tuned rather than predictive).
As \cite{Finlator_2008} show, the equilibrium metallicity of a galaxy is proportional to $1/(1 + \eta)$; therefore, a reduction of $\eta$ at high-redshifts could possibly lead to the observed increase in metallicities at $z>3$.

Regardless, it is worth emphasizing the point that the evolution in the MZR in each simulation is not consistent with any other simulation.
The closest match is Illustris and TNG, but even here the MZR evolves a factor of $2$ more in TNG than Illustris and the shape of the MZR is quite different between the two.


\subsection{FMR Predictions for the Evolution in the MZR}
\label{subsec:predicted_MZR}

\begin{figure*}
    \centering
    \includegraphics[width=0.99\linewidth]{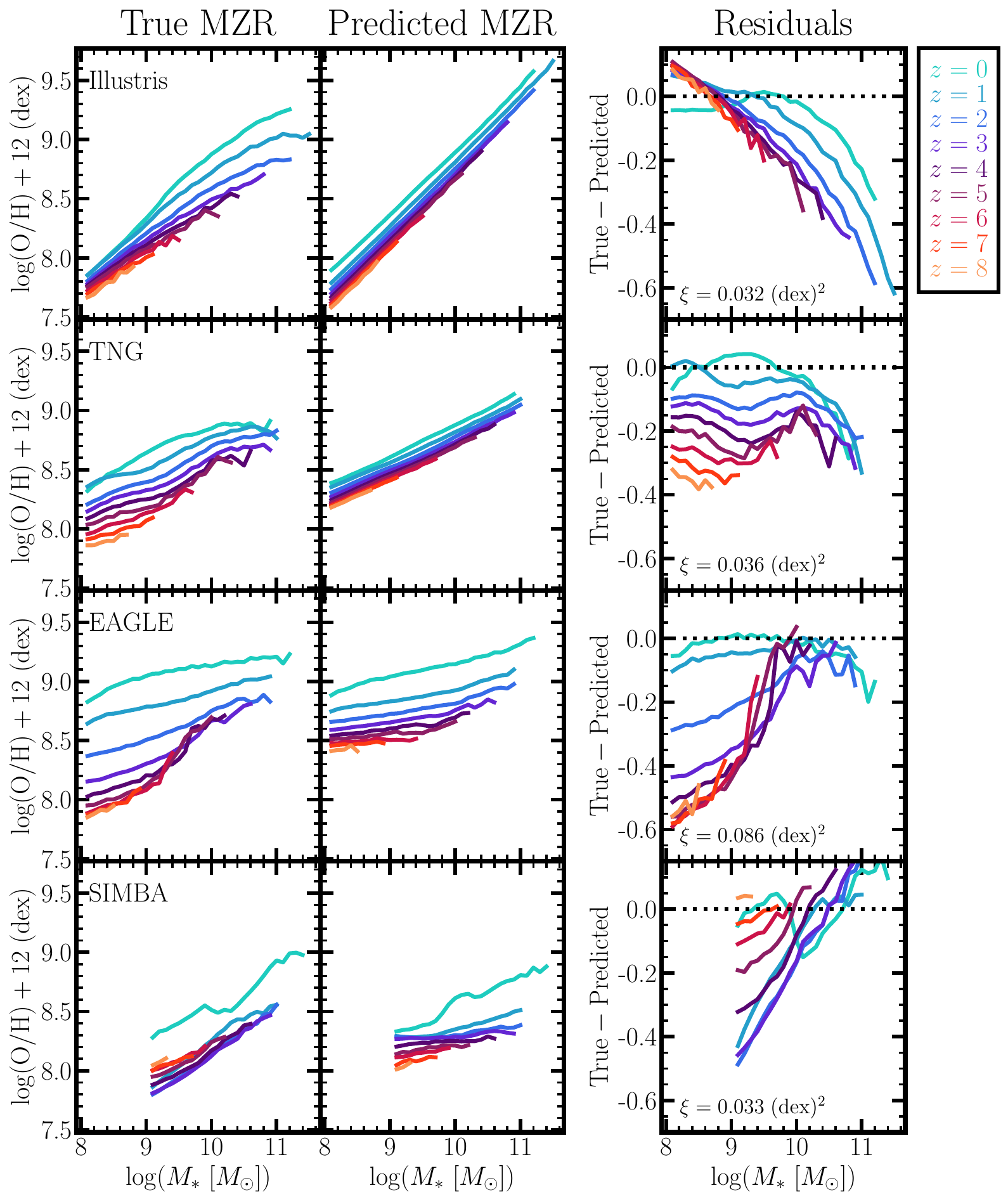}
    \caption{{\bf True median MZR versus FMR predictions of the median MZR for each simulation using a $z=0$ calibrated FMR}. {\it Left Column}: The median MZR for each redshift (same as Figure~\ref{fig:MZR_Comp}) in Illustris (top row), TNG (second row), EAGLE (third row), and SIMBA (bottom row).
    {\it Centre}: Predictions for the median MZR given a $z=0$ calibrated FMR (Equation~\ref{eqn:FMR_regressions}) and the evolution in the SFMS (Figure~\ref{fig:SFMS_Comp}).
    {\it Right}: Difference between left and central columns.
    For reference we include the mean-squared error ($\xi$) of the offsets for all the redshift bins.
    We note that the $z=0$ MZR is not a perfect reconstruction, despite the FMR being calibrated at that redshift.
    This is owing to (i) the chosen linear reconstruction of the FMR (see discussion in Appendix~\ref{appendix:different_FMRs}) and (ii) some non-zero mass dependence to the FMR (see~\protect\citeauthor{Yates_2012} \protect\citeyear{Yates_2012}, Carnevale et al. In Prep).
    }
    \label{fig:MZR_predictions}
\end{figure*}

It is worth appreciating that, for the most part, the lowest metallicities and highest SFRs are found at the highest redshifts in the simulations.
This anti-correlation between galaxies' SFR and metallicity is qualitatively consistent with the idea of the FMR.
However, it is unclear whether the similarity holds up to quantitative scrutiny.
This section therefore investigates the extent to which the combination of a $z=0$ fit FMR and SFMS can be used to predict the evolution in the MZR.

A functional form of the FMR\secondedit{, combined with the measured SFMS evolution,} allows for a straightforward way to predict high-redshift MZRs (as we discuss at length in Section~\ref{subsec:FMR_definitions}).
We demonstrate our method of predicting the MZR using the linear form for the FMR in Figure~\ref{fig:MZR_predictions}.
The left column shows the true MZR for Illustris, TNG, EAGLE, and SIMBA (top-to-bottom, respectively).
We note that the left column is identical to the results from Figure~\ref{fig:MZR_Comp} (see Section~\ref{subsubsec:MZR} for a complete discussion/comparison of the true MZRs).
The predictions are given in the central column of Figure~\ref{fig:MZR_predictions}.
We show the difference between the left and central columns in the right column of Figure~\ref{fig:MZR_predictions}, noting that some non-negligible structure exists in the $z=0$ offsets despite the FMR being calibrated at $z=0$.
The origin of the structure in the $z=0$ offsets is likely two-fold: 
(i) the MZR is inherently non-linear to some extent, which the assumed linear FMR evidently cannot capture (see discussion above and in Appendix~\ref{appendix:different_FMRs}), and 
(ii) the FMR having some non-zero mass dependence (see, e.g., \citeauthor{Yates_2012} \citeyear{Yates_2012}, \citeauthor{Alsing_2024} \citeyear{Alsing_2024}, Carnevale et al. In Preparation).

It should be noted that the predictions are more linear than the true MZRs.
This is a direct consequence of using a linear regression in fitting the FMR -- coupled to the reasonably linear nature of the SFMS.
We present these same results using a fourth-order polynomial instead in Appendix~\ref{appendix:different_FMRs}.
While we find that the shape of the predicted MZR is generally improved, the offsets at higher redshifts -- as well as some mass trends we discuss below -- persist while using a higher-order polynomial (Figure~\ref{fig:fourth_order_predictions} and Appendix~\ref{appendix:different_FMRs}).
Regardless of the specific functional form, we find a clear evolution of the MZR with respect to redshift for each simulation in both fitting methods: as redshift increases, we predict increasingly metal-poor galaxies.

As a summary metric, we provide the mean-squared error ($\xi$) of the offsets on each of the right-hand panels of Figure~\ref{fig:MZR_predictions}.
This summary metric $\xi$ is computed as the square of the residuals (true MZR $-$ predicted MZR) normalised by the number of mass bins across all redshifts.
As such, $\xi$ is robust to both 
(i) the total number of mass bins we used to create the MZR (i.e., SIMBA having a higher mass cut-off and lower redshift cut-off) as well as 
(ii) simulations having some positive but mostly negative offsets.
$\xi$ is 0.032 (dex)$^2$ for Illustris, 0.036 (dex)$^2$ for TNG, 0.086 (dex)$^2$ for EAGLE, and \secondedit{0.033} (dex)$^2$ for SIMBA.
At face value, $\xi$ tells us that the predictions in SIMBA\secondedit{, Illustris, and TNG perform equivalently well, while} EAGLE does \secondedit{worse}.

However, $\xi$ is not the entire picture.
In more detail, the offsets at higher redshift are qualitatively different in the four simulations.
The offsets have a strong mass dependence in Illustris at $z>0$.
The average metallicity is underpredicted in the lowest mass bins by $\sim0.05$ dex at each redshift, while the highest mass bins are overpredicted by as much as $0.6$ dex.
The offset for TNG's predictions actually display very little mass dependence.
Rather, there is a strong redshift dependence to the offsets with $z=1$ being on average 0.05 dex offset increasing out to $z=8$ being $\sim0.35$ dex offset.
EAGLE has both a redshift and mass dependence (though the latter is the opposite trend as Illustris).
Metallicities are underpredicted in the lowest mass bin by $0.1$ dex at $z=1$ increasing out to $0.6$ dex at $z=6-8$.
The trend with mass is such that at $z=2$, for example, the lowest mass bin is offset by $0.3$ dex while the highest mass bin is offset by $<0.1$ dex.
Finally, the average metallicity in SIMBA is overpredicted at $z\leq5$.
The offsets from the true metallicity increase to $\sim0.25$ dex at $z=3$ and then diminish at higher redshifts.
In fact, at $z=6$ the predicted MZR almost exactly matches the true MZR.
At $z=7$, however, the metallicities underpredicted.
The agreement at $z=6$ is a transient feature as the predicted values transition from being under- to over-predicted.
We therefore note that the agreement at $z=6$ is coincidental.
Furthermore, the trend of having the greatest offsets at $z=3$ with diminishing offsets at $z>3$ is qualitatively consistent with the offsets from the $z=0$ MZR seen in the previous Section (see inset on bottom right panel of Figure~\ref{fig:MZR_Comp}).
It is perhaps unsurprising that the true MZR evolution of SIMBA is not captured in this model since the SFMS is similar to those of Illustris, TNG, and EAGLE.
It is unclear how the normalisation of the MZR would increase at $z>4$ given only the $z=0$-calibrated FMR and the SFMS in SIMBA.
SIMBA also appears to have some potential trend with mass at $z=2-4$ such that low mass galaxies' metallicities are underpredicted compared to their higher mass counterparts.

\section{Discussion}
\label{sec:discussion}

\subsection{Calibrating the FMR with MZRs Across Redshift Bins}
\label{subsec:FMR_for_normalisation}

\begin{figure*}
    \centering
    \includegraphics[width=\linewidth]{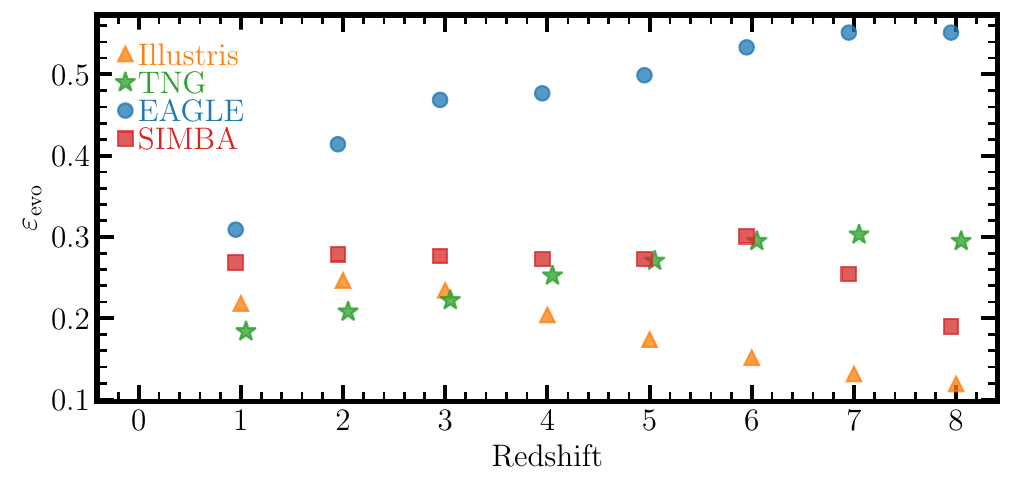}
    \caption{{\bf Evolution of $\varepsilon_{\rm evo}$ as a function of redshift.} $\varepsilon_{\rm evo}$ encodes the relative importance that SFR plays in setting the normalisation of the MZR (see Section~\ref{subsec:FMR_for_normalisation}, Equation~\ref{eqn:Chruslinkska}).
    We present $\varepsilon_{\rm evo}$ as a function of redshift in Illustris (orange triangles), TNG (green stars), EAGLE (blue circles), and SIMBA (red squares).
    See also Table~\ref{tab:alpha_table}.
    We note that the points are all offset from the integer redshifts for aesthetic reasons only.
    Evolution in $\varepsilon_{\rm evo}$ implies that the metallicity penalty galaxies pay for having higher SFRs changes with time.
    In other words, deviations in $\varepsilon_{\rm evo}$ hint at the idea that either (i) the role SFR plays in setting metallicities is a function of time (see Section~\ref{subsubsec:static_dynamic}) or (ii) additional parameters are required to describe the FMR (see Section~\ref{subsec:physics_of_FMR}).
    There is no $\varepsilon_{\rm evo}$ for $z=0$ by definition, since $\varepsilon_{\rm evo}$ is a comparative measurement to the $z=0$ relation.
    }
    \label{fig:alpha_evo}
\end{figure*}

\def\bluedagger{{\color{blue}\dagger}}
\begin{table}
    \centering
    \begin{tabular}{lx{0.15\linewidth}x{0.15\linewidth}x{0.15\linewidth}x{0.15\linewidth}}
         \toprule
          & {\bf Illustris} & {\bf TNG} & {\bf EAGLE}  & {\bf SIMBA}\\\midrule
         $z=0$ & -- & -- & -- & -- \\
         $z=1$ & $0.22$ & $0.19$ & $0.31$ & \secondedit{$0.27$}\\
         $z=2$ & $0.25$ & $0.21$ & $0.41$ & \secondedit{$0.28$}\\
         $z=3$ & $0.23$ & $0.23$ & $0.47$ & \secondedit{$0.28$}\\
         $z=4$ & $0.20$ & $0.25$ & $0.47$ & \secondedit{$0.26$}\\
         $z=5$ & $0.17$ & $0.27$ & $0.50$ & \secondedit{$0.29$}\\
         $z=6$ & $0.15$ & $0.30$ & $0.54$ & \secondedit{$0.31$}\\
         $z=7$ & $0.13$ & $0.30$ & $0.56$ & \secondedit{$0.25$}\\
         $z=8$ & $0.11$ & $0.30$ & $0.55$ & \secondedit{$0.19$}\\\bottomrule
    \end{tabular}
    \caption{{\bf All $\varepsilon_{\rm evo}$ values for Illustris, TNG, EAGLE, and SIMBA.}
    $\varepsilon_{\rm evo}$ (see Equation~\ref{eqn:Chruslinkska}) is the parameter that relates how important changes in average SFRs are in setting the average metallicity of galaxies.
    We show these values in Figure~\ref{fig:alpha_evo}.
    }
    \label{tab:alpha_table}
\end{table}

In the previous sections, we found that the combination of the $z=0$-calibrated FMR and evolution in the SFMS is insufficient in describing the normalisation evolution of the MZR in Illustris, TNG, EAGLE, and SIMBA.
This finding is potentially indicative of two things: 
(i) a change in the importance SFR plays in setting the normalisation or 
(ii) additional parameter dependencies required in setting the normalisation of the MZR.
In this section, we investigate the former of these two implications.
We discuss the latter in Section~\ref{subsec:physics_of_FMR}.

Recall that Equation~\ref{eqn:MZR_full} is an approximation to the MZR at any redshift (where $m,b,~{\rm and}~\alpha$ are determined at $z=0$).
We can take the difference of $\langle Z(M_*,z=i)\rangle$ and $\langle Z(M_*,z=0)\rangle$.
Critically, a redshift-invariant FMR requires that the parameters of the fit ($m, b, \alpha$) do not change with redshift.
Therefore this difference can be written as
\begin{equation}
    \label{eqn:FMR_penalty}
    \langle Z(M_*, z=i) \rangle - \langle Z(M_*, z=0) \rangle = - m \alpha \log \left( \frac{\langle {\rm SFR} (M_*, z=i) \rangle}{\langle {\rm SFR} (M_*, z=0) \rangle} \right) ~.
\end{equation}
This equation states the changes observed in the evolution of the MZR is set by the changes in the average SFR, scaled by some ``penalty'' term, $-m\alpha$.
Furthermore, Equation~\ref{eqn:FMR_penalty} is a statement that increased SFRs are the key to describing decreased metallicities at higher redshifts ($z_i$, where $i$ is any redshift other than 0) via a comparison to that of $z=0$.
The penalty weighting ($-m\alpha$) is such that galaxies with higher-than-average SFRs have lower metallicities (and vice versa).
We caution that this formulation is inconsistent with previous results: $\alpha$ has been shown to vary significantly as a function of redshift in each of these simulations (see \citetalias{Garcia_2024b} for Illustris, TNG and EAGLE, Appendix~\ref{appendix:SIMBA_strong_weak} for SIMBA).
Regardless, this form of the FMR is useful phenomenologically if we rewrite this slightly such that
\begin{equation}
    \label{eqn:Chruslinkska}
    \Delta [{\rm MZR}(M_*)]_{z=0\to i} = -\varepsilon_{\rm evo} ~\Delta [\log{\rm SFMS}(M_*)]_{z=0\to i}~,
\end{equation}
where $\varepsilon_{\rm evo}$ is a more generalised penalty term and the subscript $z=0\to i$ implies the change from $z=0$ to $z=i$.\ignorespaces
\footnote{
We note that Equation~\ref{eqn:Chruslinkska} is essentially a derivation of Equation~2 of \cite{Chruslinska_2021}.
However, \cite{Chruslinska_2021} present this for the scatter about the MZR, whereas we characterise the normalisation of the MZR.
}.
In detail, $\varepsilon$ should be roughly $m\alpha$.
Yet, we gain flexibility in the value of this ``penalty'' by removing the restriction that it is strictly tied to the the slope of the FMR and scatter about the MZR.
Physically, $\varepsilon$ encodes how important changes in the average SFR are in setting the average metallicity.
If $\varepsilon$ is higher, a change in SFRs will correspond to a larger penalty in metallicty. 
In addition to standardizing our companions across simulations (which may use different values of both $\alpha$ and $m$ independently), summarizing the redshift evolution with the $\varepsilon_{\rm evo}$ penalty term can normalize comparisons with observations.

Equation~\ref{eqn:Chruslinkska} sets up an experiment in which we can determine the $\varepsilon_{\rm evo}$ value that minimises the offset (i.e., mean-squared error $\xi$) in the predictions of the high-redshift MZR.
We vary $\varepsilon$ from 0 to 1 in steps of 0.01 to produce predicted MZRs at each redshift in each simulation\ignorespaces
\footnote{
There is no physical reason the value of $\varepsilon_{\rm evo}$ could not be $>1$; however, in our testing we find that it is always less than $1$.
}.
As before, we quantify the difference between the predicted MZR and the true MZR using $\xi$; however, we compute $\xi$ at each redshift individually, not in aggregate for the $\varepsilon$ determination.
The $\varepsilon$ value that minimises $\xi$ is chosen and is henceforth referred to as $\varepsilon_{\rm evo}$. 
We define an uncertainty on $\varepsilon_{\rm evo}$ via a bootstrap analysis.
We resample the galaxies at each redshift with replacement and determine $\varepsilon_{\rm evo}$ 1,000 times.
We find that the value of $\varepsilon_{\rm evo}$ is usually within $\pm$0.01 from the values reported in Table~\ref{tab:alpha_table}.

Figure~\ref{fig:alpha_evo} presents the $\varepsilon_{\rm evo}$ values for Illustris as orange triangles, TNG as green stars, EAGLE as blue circles, and SIMBA as red squares (values also listed in Table~\ref{tab:alpha_table}).
We note that constructing the FMR in this way is comparative.
The calibration process at each redshift is done with respect to the $z=0$ average masses, metallicities and SFRs.
$\varepsilon_{\rm evo}$ at $z=0$ is therefore undefined.
We find that each simulation has a non-zero $\varepsilon_{\rm evo}$ value at each redshift, suggesting that SFR plays some role in setting the normalisation of the MZR at each redshift.
In Illustris, we find that $\varepsilon_{\rm evo}$ has only a very weak trend with redshift, increasing at $z=2-3$ and then decreasing at $z\geq4$.
In TNG, we find that $\varepsilon_{\rm evo}$ is around $0.15$ at $z=1$ and increases nearly monotonically to $0.30$ at $z=8$.
$\varepsilon_{\rm evo}$ is increases at $z=1-3$ in EAGLE, is roughly constant at $\sim\!0.47$ from $z=3=5$, and jumps to a value of $\sim\!0.55$ at $z=6-8$.
SIMBA's $\varepsilon_{\rm evo}$ is roughly constant at \secondedit{$\sim\!0.27$} until $z=6$ \secondedit{where it rises slightly}, but then decreases at \secondedit{$z=7~{\rm and}~8$.
}

\subsubsection{What do variations in \texorpdfstring{$\varepsilon_{\rm evo}$}{epsilon evo} mean?}
\label{subsubsec:variations_alpha_evo}

\begin{figure*}
    \centering
    \includegraphics[width=0.99\linewidth]{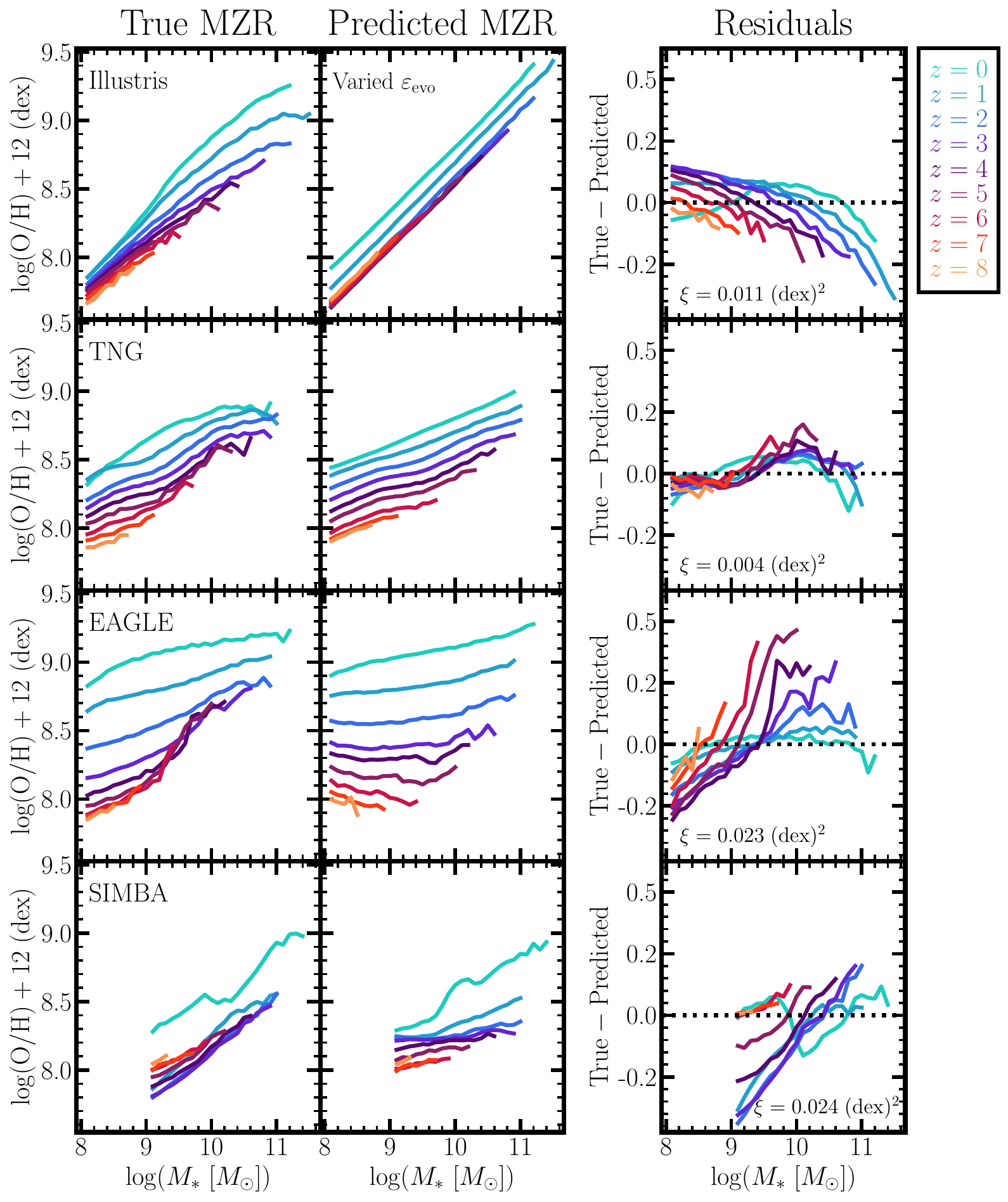}
    \caption{{\bf Predictions for the median MZR using a variable $\varepsilon_{\rm evo}$}. Same as Figure~\ref{fig:MZR_predictions} but instead of using a $z=0$ calibrated FMR (i.e., using Equation~\ref{eqn:MZR_full}) we now allow the role SFR plays to vary as a function of redshift using Equation~\ref{eqn:Chruslinkska}.
    For comparison, we include the mean-squared error ($\xi$) of all offsets across all the redshift and mass bins in the right column panels.
    }
    \label{fig:varied_epsilon_predictions}
\end{figure*}

While phenomenological, Equation~\ref{eqn:Chruslinkska} represents a reasonable interpretation of the main concept of the FMR: as SFRs decrease, metallicities increase.
In this section, we anchor this phenomenological treatment analytically as well as explore what variations in $\varepsilon_{\rm evo}$ mean for our predictions of the normalisation of the MZR.

We present the variable $\varepsilon_{\rm evo}$ predictions for the high-$z$ MZR in Figure~\ref{fig:varied_epsilon_predictions} to quantify these changes in the penalty term (Illustris in top row, TNG in second row, EAGLE in third row, and SIMBA in fourth row).
We note that the left-hand column of Figure~\ref{fig:varied_epsilon_predictions} is identical to both Figure~\ref{fig:MZR_Comp} as well as the left-hand column of Figure~\ref{fig:MZR_predictions}.
The central column of Figure~\ref{fig:varied_epsilon_predictions} is qualitatively similar to the central column of Figure~\ref{fig:MZR_predictions}; however, instead of using the $z=0$ calibrated FMR, these predictions are made using the varied $\varepsilon_{\rm evo}$\ignorespaces
\footnote{
In order to make a fair comparison to the methodology of Section~\ref{sec:results}, we define the $z=0$ MZR by ``reconstructing'' it via the linear FMR (i.e., Equation~\ref{eqn:FMR_regressions}).
We note that this is different than the methodology used to determine $\varepsilon_{\rm evo}$.
The only difference between using a reconstructed MZR and the true MZR is the shape of the predictions (similar to different functional forms, as in Appendix~\ref{appendix:different_FMRs}).
}.
The shapes of the predicted MZRs are more linear than their true MZR counterparts (similar to Section~\ref{subsec:predicted_MZR}).
This is in large part owing to the chosen reconstruction of the $z=0$ MZR using the linear FMR.
As we show in Appendix~\ref{appendix:different_FMRs} for the $z=0$ FMR and SFMS combination predictions, the shape of the MZR is improved by using a higher-order polynomial.
The qualitative trends, however, remain.
We therefore choose this reconstruction of the $z=0$ MZR as the most fair comparison with the results presented in Section~\ref{subsec:predicted_MZR}.
Regardless of choice in $z=0$ MZR reconstruction, we find that fitting $\varepsilon_{\rm evo}$ to each redshift individually reconstructs the evolution in the normalisation of the MZR much more faithfully than the $z=0$ FMR and SFMS combination of Section~\ref{subsec:predicted_MZR}.
In fact, by using a variable $\varepsilon_{\rm evo}$, we can now reproduce the MZR ``turn around'' in SIMBA.

The right-hand column of Figure~\ref{fig:varied_epsilon_predictions} shows the offsets between the true MZRs (left column) from that of the predicted MZRs (central column).
As before, we quantify these offsets using the mean-squared error, $\xi$ (presented for each simulation on the right-hand panels).
We find the $\xi$ is 0.010 (dex)$^2$ for Illustris, 0.003 (dex)$^2$ for TNG, 0.023 (dex)$^2$ for EAGLE, and \secondedit{0.024} (dex)$^2$ for SIMBA.
It should be noted that $\xi$ has decreased in each simulation compared to those of the $z=0$ FMR prediction (see Figure~\ref{fig:MZR_predictions}).
The decrease in $\xi$ is perhaps unsurprising given the methodology of determining $\varepsilon_{\rm evo}$ specifically minimises $\xi$.
What may be surprising is the magnitude of the decrease.
In Illustris $\xi$ is decreased by a factor of $3.2$, for TNG $\xi$ decreases by a factor of $12$, for EAGLE $\xi$ decreases by a factor of $3.7$, and for SIMBA $\xi$ decreases by a factor of \secondedit{$1.4$}.
It is therefore clear that using a variable $\varepsilon_{\rm evo}$ does a significantly better job quantifying the evolution in the normalisation of the MZR than the $z=0$ FMR and SFMS combination.
We do note that the mass trends seen previously in the offsets (right column of Figure~\ref{fig:MZR_predictions}) persist, however, when using a variable $\varepsilon_{\rm evo}$.

It is worth mentioning that all of the analysis above follows the assumption that what drives the evolution of the normalisation of the MZR is the evolution in SFRs alone.
While we have some success reproducing the evolution of the MZR using a variable $\varepsilon_{\rm evo}$ (e.g., Figure~\ref{fig:varied_epsilon_predictions}), it should be noted that additional galactic parameters (e.g., gas masses, gas fractions, galaxy sizes, etc) will play a role in setting the metallicity of galaxies.
It is therefore possible that using SFRs alone to reproduce MZR evolution in this way will neglect other contribution to galaxies' metal evolution.
We discuss this idea further in Section~\ref{subsec:physics_of_FMR}.

\subsection{Static versus Dynamic FMR}
\label{subsubsec:static_dynamic}

In \citetalias{Garcia_2024b}, we investigated whether or not $\alpha_{\rm min}$ -- a parameter encoding the importance of SFR in setting the scatter of the MZR -- varies as a function of redshift.
We introduced the ``strong'' and ``weak'' FMRs to describe whether or not $\alpha_{\rm min}$ had a redshift dependence (with strong meaning no redshift dependence and weak indicating some redshift dependence).
We extend the analogy to the FMR for scatter made in the previous section further by introducing the ``static'' and ``dynamic'' FMRs to describe the FMR for normalisation.
The static FMR describes a relation in which $\varepsilon_{\rm evo}$ is constant as a function of redshift, whereas the dynamic FMR is where $\varepsilon_{\rm evo}$ varies with redshift.
Put another way, the schematic presented in Figure~\ref{fig:cartoon} implicitly implies a static FMR; wherein the combination of the SFMS and $z=0$ FMR can describe the $z>0$ MZR.

Critically, while related, a strong FMR does not necessarily imply a static FMR, nor does a dynamic FMR imply a weak FMR.
The scatter about the MZR may or may not be agnostic to the changes in the overall normalisation (or vice versa).
The critical advantage of building the FMR framework in this way is to develop the language for describing the two features of interest in the MZR: scatter and evolution in the normalisation -- both of which nominally contribute to the classification of the FMR as fundamental.
We discuss this more in Section~\ref{subsec:taking_stock}.


We perform a \secondedit{Mann-Kendall trend test \citep{Mann_1945,Kendall_1975,Hussain_2019}} to more concretely classify the average FMRs as static or dynamic.
\secondedit{A Mann-Kendall test is a robust non-parametric test that specialises in detecting trends in time-series data.
The advantage of this test is not only to determine whether a statistically significant trend exists, but what its directionality is (i.e., increasing/decreasing trend)\ignorespaces
\footnote{\ignorespaces
\secondedit{\ignorespaces
It should be noted that the Mann-Kendall test is not robust to oscillatory data. 
The test would likely not be able to detect a trend if $\varepsilon_{\rm evo}$ were changing in some sort of periodic pattern (i.e., increasing and then decreasing).
However, by visual inspection of Figure~\ref{fig:alpha_evo}, none of the simulations exhibit oscillatory behaviour that might call into question the efficacy of the Mann-Kendall test.
}}.
}
The null hypothesis \secondedit{of this test} is that \secondedit{there is no trend in $\varepsilon_{\rm evo}$ with time.}
We classify a simulation\secondedit{'s} FMR as static if the null hypothesis cannot be rejected.
Conversely, if we reject the null hypothesis, the simulations' FMR for normalisation is classified as dynamic.
\secondedit{The directionality of the Mann-Kendall test also allows us to define an ``increasing dynamic'' and ``decreasing dynamic'' FMR.}
\secondedit{We find $s$-statistics of $-24.0$ in Illustris, $26.0$ in TNG, $26.0$ in EAGLE, and $6.0$ in SIMBA.
These correspond to $p$-values of $4.43\times10^{-3}$ for Illustris, $1.98\times10^{-3}$ for TNG, $1.98\times10^{-3}$ for EAGLE, and $0.53$ in SIMBA.
We reject the null hypothesis for Illustris, TNG and EAGLE at the 0.05 confidence level, but cannot reject it for SIMBA.
We therefore classify SIMBA static and Illustris, TNG and EAGLE as dynamic.
Moreover, based on the sign of the $s$-statistic, we can specify Illustris as ``decreasing dynamic'' and TNG/EAGLE as ``increasing dynamic''.
}

\section{Does the Fundamental Metallicity Relation Evolve with Redshift?}
\label{subsec:taking_stock}

\begin{figure*}
    \centering
    \includegraphics[width=\linewidth]{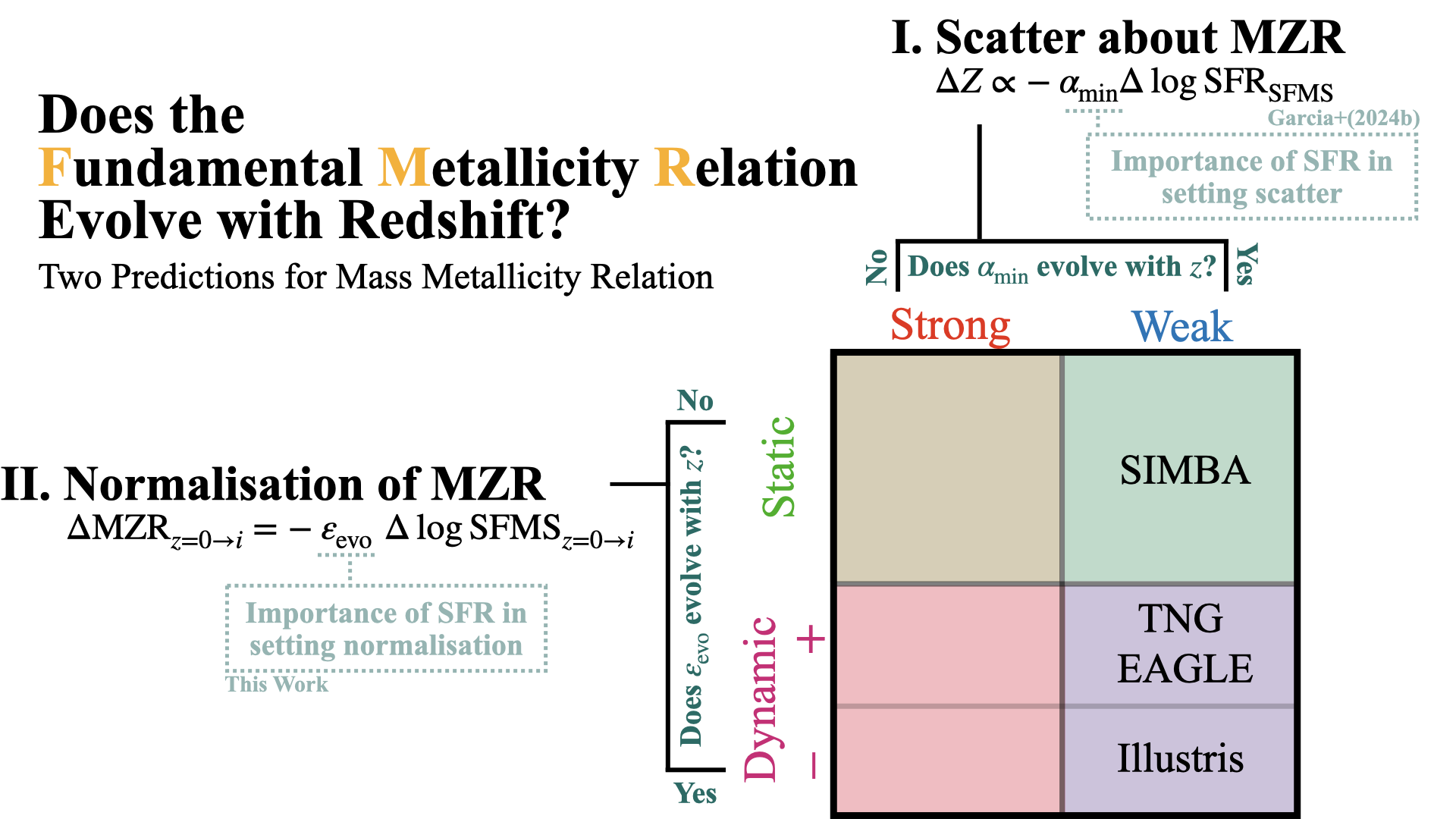}
    \caption{{\bf Does the Fundamental Metallicity Relation Evolve With Redshift?}
    Illustration of the findings of this work and \protect\citeauthor{Garcia_2024b} (\protect\citeyear{Garcia_2024b}, henceforth \protect\citetalias{Garcia_2024b}).
    The FMR makes predictions for two features of the MZR: (i) its scatter (discussed in \protect\citetalias{Garcia_2024b}) and (ii) the evolution in its normalisation (this work).
    The scatter about the MZR ($\Delta Z$) is proportional to minus $\alpha_{\rm min}$ -- the scatter penalty parameter -- times the \secondedit{SFRs' offset from the SFMS}.
    The normalisation evolution ($\Delta {\rm MZR}$), on the other hand, is equal to minus $\varepsilon_{\rm evo}$ -- the evolutionary penalty parameter -- times the change in the SFMS ($\Delta {\rm SFMS}$).
    Both proportionalities are such that an increase in SFR implies a decrease in metallicity.
    The grid in the lower right-hand corner shows the four different classifications possible based on $\alpha_{\rm min}$ and $\varepsilon_{\rm evo}$.
    If the scatter penalty parameter, $\alpha_{\rm min}$, is fixed with redshift the FMR is ``strong'', otherwise it is ``weak'' (see \protect\citetalias{Garcia_2024b}).
    If the evolutionary penalty parameter, $\varepsilon_{\rm evo}$, is fixed with redshift the FMR is ``static'', otherwise it is ``dynamic'' (see Section~\ref{subsubsec:static_dynamic} of this work).
    \secondedit{We further refine ``dynamic'' into increasing ($+$) and decreasing ($-$) based on the Mann-Kendall trend test.
    We find that the FMR in SIMBA is both static and weak, the FMRs in TNG and EAGLE are increasing dynamic and weak, and the FMR in Illustris is decreasing dynamic and weak.}
    We further speculate that recent observed high-redshift offsets from the FMR may indicate that the FMR in observations is \secondedit{increasing} dynamic (see Section~\ref{subsec:JWST_implications}); however, further testing is required to confirm this.
    }
    \label{fig:summary}
\end{figure*}

In light of this work and \citetalias{Garcia_2024b}, it is worth contextualising the landscape of the FMR and directly addressing the title question of these investigations: ``Does the FMR evolve with Redshift?''.
We summarise the findings of both works in Figure~\ref{fig:summary}.

The FMR makes two key predictions:
(i) scatter about MZR is correlated with SFRs and 
(ii) the normalisation evolution of the MZR itself is correlated with SFRs.
In these works, we have attempted to address FMR evolution by understanding each of these individual features separately.
We identify a ``SFR penalty'' term for both features, $\alpha_{\rm min}$ for scatter and $\varepsilon_{\rm evo}$ for normalisation.
For the scatter about the FMR, we define a Strong FMR as one where $\alpha_{\rm min}$ does not evolve and a Weak FMR where $\alpha_{\rm min}$ {\it does} evolve.
For the normalisation evolution of the FMR, we define a Static FMR as one where $\varepsilon_{\rm evo}$ does not evolve and a Dynamic FMR as one where $\varepsilon_{\rm evo}$ {\it does} evolve.
We find that SIMBA has a Static-Weak FMR -- the role SFR plays in the scatter about the MZR changes with time, but not the normalisation -- and that \secondedit{Illustris}, TNG, and EAGLE have a Dynamic-Weak FMR -- the role SFR plays in both the scatter and normalisation of the MZR changes with time.

By defining an evolving FMR as one in which either $\alpha_{\rm min}$ or $\varepsilon_{\rm evo}$ (or both) have redshift evolution, the FMR evolves with redshift in each of the four simulations analysed here.
In this section, we explore the potential implications for recent high-redshift \JWST{} observations as well as the utility of the FMR as a diagnostic for galaxy evolution physics.

\subsection{Implications for Recent JWST Observations at high redshift}\label{subsec:JWST_implications}

\begin{figure}
    \centering
    \includegraphics[width=0.99\linewidth]{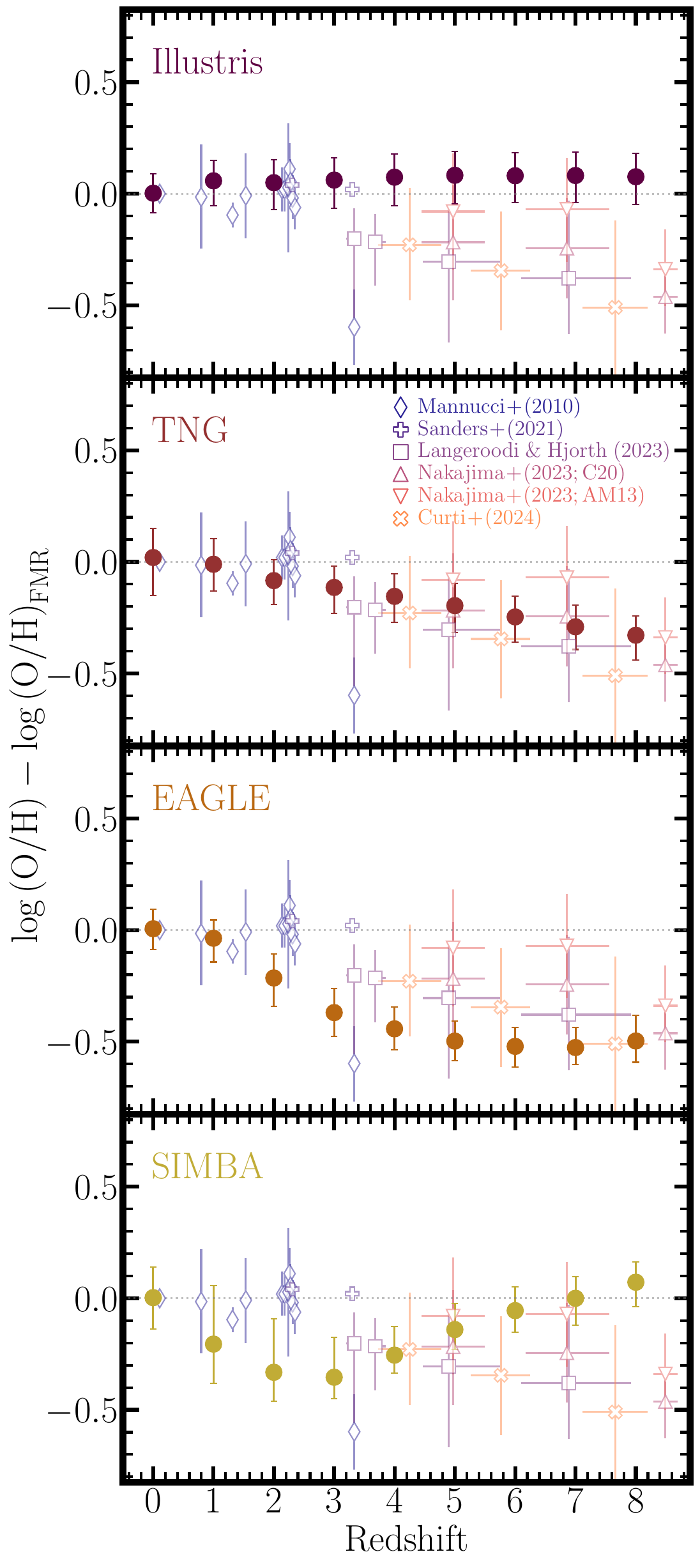}
    \caption{{\bf Redshift evolution of offsets from the FMR.} Offsets from the FMR at all redshifts based on the $z=0$ fit. The filled circles are the distribution of offsets from the simulation in each panel (top-to-bottom: Illustris, TNG, EAGLE, and SIMBA) at each redshift. 
    The error bars on the simulation data points are the $16^{\rm th}$ and $84^{\rm th}$ percentiles of the distribution of offsets. Observational data of offsets from the FMR are shown in the hollow diamonds from \protect\cite{Mannucci_2010}, plus signs from \protect\cite{Sanders_2021}, Xs from \protect\cite{Curti_2023}, squares from \protect\cite{Langeroodi_2023}, and triangles from \protect\cite{Nakajima_2023}. The difference between the \protect\cite{Nakajima_2023} triangles is the underlying FMR compared against: pointing up compares to \citeauthor{Curti_2020} (\citeyear{Curti_2020}; $\alpha_{\rm min}=0.55$) and pointing down compares to \citeauthor{Andrews_Martini_2013} (\citeyear{Andrews_Martini_2013}; $\alpha_{\rm min}=0.66$). }
    \label{fig:offsets}
\end{figure}

Recent JWST observations of the early universe ($z\gtrsim4$) have found that the metallicity predicted by the FMR is typically much higher than observed \citep[e.g.,][]{Heintz_2022,Langeroodi_2023,Nakajima_2023,Castellano_2024,Curti_2023}.
These observational results are in qualitative agreement with the model proposed in Section~\ref{subsec:predicted_MZR}, where we find that the $z=0$ calibrated FMR has difficulties predicting high-redshift MZRs.
Here, we make a more direct comparison between the observed high-redshift offsets and those of simulations.

We compute offsets from the FMR for each individual galaxy using the same methodology described in Section~\ref{subsec:FMR_definitions} and used in Section~\ref{subsec:predicted_MZR}.
The metallicity of a galaxy is predicted by plugging its mass and SFR into the $z=0$ FMR (Equation~\ref{eqn:FMR_regressions}).
We follow the convention set by observations of negative offsets implying that a galaxy is metal poor compared to FMR predictions.
We show the median offsets from the $z=0$ FMR for each simulation in Figure~\ref{fig:offsets} (Illustris, TNG, EAGLE, and SIMBA are top-to-bottom, respectively).
The errorbars on the simulation data points are the $16^{\rm th}$ and $84^{\rm th}$ percentiles of offsets.
We note that all of the residuals are centered around zero at $z=0$ by construction.
We find that Illustris' offsets have a slight trend of more positive offsets with increasing redshift.
It should be noted, however, that the width of the distribution of offsets includes no offsets at all redshifts in Illustris.
In TNG and EAGLE, the offsets are systematically offset from zero with increasing redshift.
More specifically, the offsets from the $z=0$ calibrated FMR become more negative roughly linearly with increasing redshift in TNG.
The offsets become more negative with increasing redshift in EAGLE out to $z\sim5$ and then plateau at $z=6-8$.
The offsets from the $z=0$ FMR in SIMBA become more negative out to $z=3$, decrease to no offsets at $z=5$, and then become positive at \secondedit{$z=8$}.
The behaviour of the offsets from the FMR in each simulation is entirely consistent with the results presented in Section~\ref{subsec:predicted_MZR} (Figure~\ref{fig:MZR_predictions}).

As a point of comparison, we overplot summary statistics of offsets from the FMR at $z=0-8+$ in observations from \citeauthor{Mannucci_2010} (\citeyear{Mannucci_2010}; open diamonds), \citeauthor{Sanders_2021} (\citeyear{Sanders_2021}; open plus signs), \citeauthor{Langeroodi_2023} (\citeyear{Langeroodi_2023}; open squares), \citeauthor{Nakajima_2023} (\citeyear{Nakajima_2023}; open triangles\ignorespaces
\footnote{
\cite{Nakajima_2023} compare to both \citeauthor{Curti_2020} (\citeyear{Curti_2020}; triangles pointing up) and \citeauthor{Andrews_Martini_2013} (\citeyear{Andrews_Martini_2013}; triangles pointing down). Although \citeauthor{Nakajima_2023} prefer the comparison to \citeauthor{Andrews_Martini_2013} as \citeauthor{Andrews_Martini_2013} probed a wider $\mu_{\alpha}$ space.
We include comparisons to both as note that different $\alpha_{\rm min}$ values can impact the magnitude of offsets.
}\ignorespaces
), and \citeauthor{Curti_2023} (\citeyear{Curti_2023}; open Xs) in Figure~\ref{fig:offsets}.
\edit{We note that each of these observational works rely on different methodology (including derived metallicities, see further discussion below) for obtaining offsets from the FMR.
We therefore caution that some uncertainty exists in the scatter of these points based on different diagnostics and sample selection; however, we include this data as an observational counterpoint to our simulation results in previous sections.}
The magnitude of the offsets in observations is, on aggregate, larger than those we find in Illustris and SIMBA but comparable to that of TNG and EAGLE.
Strikingly, the negative offsets of TNG and EAGLE are within the errorbars of \cite{Langeroodi_2023}, \citeauthor{Nakajima_2023} (\citeyear{Nakajima_2023}, comparisons to \citeauthor{Curti_2020} \citeyear{Curti_2020}), and \cite{Curti_2023}.
It should be noted, however, that \citeauthor{Nakajima_2023} (\citeyear{Nakajima_2023}, comparisons to \citeauthor{Andrews_Martini_2013} \citeyear{Andrews_Martini_2013}) do not show significant offsets until past $z\sim8$.
This evolution occurs much later than in TNG and EAGLE, yet could be roughly consistent with Illustris and SIMBA which both have similar offsets in that redshift range.
We note that there are two challenges in these comparisons:
(i) the derivation of metallicities (particularly at high-$z$) is not straight-forward in observations and
(ii) sample sizes at high-redshift, both in simulations and observations, are rather limited.
For both of these reasons (detailed more below) we caution against too strong of an interpretation of the comparison with observational results.

Briefly, the determination of gas-phase metallicities in observations relies on emission line spectra from, e.g., [O~{\sc iii}]  for the direct method \citep[see][and references therein]{Kewley_2019,Maiolino_Mannucci_2019}.
The [O~{\sc iii}] $\lambda$4363 emission line was virtually unattainable at $z\gtrsim3$ (the emission line is redshifted out of the spectral range of most optical instruments) until the recent deployment of NIRSpec aboard \JWST.
Having the [O~{\sc iii}] $\lambda$4363 line is critical since it is used to calibrate so-called ``strong-line diagnostics'' \citep[e.g.,][]{Curti_2017,Sanders_2021,Nakajima_2022}\ignorespaces
\footnote{
We note that some strong-line diagnostics are also sometimes calibrated using photoionisation models \citep[e.g.,][]{Kewley_2002,Perez-Montero_2014}.
}.
Strong-line diagnostics are what population studies are typically built upon \citep[e.g.,][]{Tremonti_2004,Ellison_2008, Mannucci_2010} since obtaining metallicities based on the direct method can be difficult owing to an oftentimes faint [O~{\sc iii}] $\lambda$4363 line.
However, even at low-$z$, different strong line calibrators can disagree on a metallicity by 0.7 dex (\citeauthor{Kewley_Ellison_2008} \citeyear{Kewley_Ellison_2008}).
Much work has therefore gone into re-calibrating strong-line relations for use at high redshift due to the evolution in ISM ionization conditions \citep[e.g.,][]{Curti_2023a,Sanders_2023,Trump_2023,Ubler_2023,Laseter_2024} including using cosmological simulations \citep{Garg_2023, Hirschmann_2023}.
As such, the derived metallicities of the \JWST{} works may be subject to change depending on future constraints of metallicity diagnostics.
Beyond deriving even-handed metallicity measurements, there is a question of sample size and completeness at high-$z$.
The galaxy populations at low redshift are very well-sampled thanks to large surveys like the Sloan Digit Sky Survey (SDSS; \citeauthor{Abazajian_2009} \citeyear{Abazajian_2009}).
At higher redshift, however, galaxy populations are much more coarsely sampled.
To put this concretely, recent high-redshift \JWST{} observations use $\lesssim400$ galaxies across $z=3-10$ \citep{Langeroodi_2022,Nakajima_2023,Curti_2023}.
We also note that simulations suffer from similarly sparse statistics at higher redshifts.
With increased sample sizes -- combined with more accurate metallicity diagnostics -- the redshift evolution of the FMR in observations can more faithfully be assessed.
\secondedit{In addition, there may be some inherent differences between the observational and simulation samples (e.g., no attempt was made to mass match the samples).
As such, there could be some inherent mass dependencies either missing from the observed sample or inherent to our simulated sample (see mass trends from right-hand panels of Figure~\ref{fig:MZR_predictions}) that are left unaccounted for in this comparison.
}

With all of that being said, it is interesting to note that the agreement between simulations and the bulk of the recent observations comes between TNG and EAGLE -- which both have \secondedit{increasing} dynamic FMRs (see Section~\ref{subsubsec:static_dynamic}).
If these high redshift observations are robust to the aforementioned issues, it is likely that the high-redshift FMR offsets indicate that the observed FMR may be \secondedit{increasing} dynamic as well.
This agreement in behaviour between EAGLE, TNG, and observations is certainly suggestive; however, an explicit test of $\varepsilon_{\rm evo}$ evolution is required in observations to confirm whether the FMR is truly dynamic as we have defined it here.
\secondedit{Regardless, it is important to note that, despite each of the simulations exhibiting a ``weak'' FMR, the behaviours in the offsets at high-redshift are {\it completely} different.}

\subsection{The FMR as a Diagnostic Tool for the Baryon Cycle}
\label{subsec:physics_of_FMR}

As discussed in Section~\ref{sec:intro}, the FMR arises naturally from the baryon cycle. 
Competition between pristine gas accretion from the CGM and metal enrichment from the ISM drives perturbations from the MZR \citep{Torrey_2018}.
The idea of a redshift-invariant FMR -- either ``static'' or ``strong'' -- insists that the relative role that each component of the baryon cycle is not a function of time.
As we have shown in this paper, as well as \citetalias{Garcia_2024b}, this is not the case for any of the simulations analysed here.
By assuming the FMR is invariant, we are marginalising over salient information about the evolution of galaxies.
Yet, the baryon cycle is treated very differently between each simulation model \citep{Wright_2024}.
Since the FMR is effectively an observational probe of the baryon cycle, it is perhaps unsurprising that there is some level of evolution in the FMR.

The FMR is a result of the complex interplay of the baryon cycle; therefore, determining what exactly sets the FMR is difficult.
The issue becomes more complex as the FMR attempts to simplify these complex processes by using a single diagnostic for the current state of the baryon cycle: SFR.
While a useful proxy, condensing the entire picture of the baryon cycle into the current SFR appears to be too broad a generalisation.
Acknowledging the limitations of a single, static FMR and instead considering a more generalized relation that may have significant redshift evolution opens a very rich area of study for understanding the baryon cycle of observations and constraining the baryon cycle of simulations.

There are a number of mechanisms that are simply not taken into account in the FMR that may be driving its evolution.
To name a few:
(i) gas in- and out- flows, 
(ii) galactic winds,
(iii) stellar and AGN feedback,
(iv) mergers, and
(v) environment.
A detailed investigation of each of these effects on the FMR is beyond the scope of the present work.
However, we briefly summarise literature below which investigates the effects of each mechanism on either the FMR and/or the metal content of galaxies.

Analytic models show that the combination of gas inflows, outflows, and recylcing are what drive SFRs \citep[see][etc]{Finlator_2008,Dave_2012}.
\cite{Wright_2024} compare the gas flow rates of TNG, EAGLE, and SIMBA and find that each simulation has quantitatively different behaviours.
The differences in the implementation of the gas flows may therefore manifest themselves in the differences between each simulation's FMR behaviour (e.g., Equation~\ref{eqn:FMR_regressions}, Figure~\ref{fig:alpha_evo}).
Interestingly, \citeauthor{Bassini_2024} (\citeyear{Bassini_2024}), show that gas fractions/SFRs are not what drives the evolution of the MZR in the FIRE model (using the FIREBox simulations; \citeauthor{Feldmann_2023} \citeyear{Feldmann_2023}).
Rather, \cite{Bassini_2024} attribute the origin of the normalisation of the MZR to galactic inflows and outflows.

The normalisation of the MZR appears similarly sensitive to the mass loading factors ($\eta$) of galactic winds. 
This was shown to be the case in \cite{Finlator_2008}, who demonstrate that the equilibrium metallicity of galaxies scales as $1/(1+\eta)$.
Increased mass loading of the winds should directly correspond to a decrease in the equilibrium metallicity of galaxies.
As an example of the impact of galactic winds, we suggest that the artificially decreased mass loading factors at high-redshift in SIMBA are what cause a ``turn-around'' in the normalisation evolution at $z>3$.
The FMR, and the potential redshift evolution thereof, can therefore provide a discriminator between different galactic wind prescriptions.

Despite the discrepancies between the gas flows in the models analysed here, a striking commonality is that they all have some sub-grid implementation of the star-forming ISM (\citeauthor{Springel_Hernquist_2003} \citeyear{Springel_Hernquist_2003} in Illustris and TNG, \citeauthor{Schaye_DallaVechhia_2008} \citeyear{Schaye_DallaVechhia_2008} in EAGLE, and \citeauthor{Krumholz_2011} \citeyear{Krumholz_2011} in SIMBA).
Yet, models with a more explicit treatment of the ISM that more directly resolve the sites of star formation exist (see, e.g., Feedback In Realistic Environments, FIRE, simulations; \citeauthor{Hopkins_2014} \citeyear{Hopkins_2014}).
The choice of ISM model may have a significant impact on the derived MZR evolutionary properties.
The explicit ISM model of FIRE has burstier stellar feedback -- more feedback over a short timescale.
It is possible that the lack of importance found in SFRs/gas fractions in \cite{Bassini_2024} stems from the different feedback implementation in FIRE.
Bursty feedback my curtail the role that SFRs play (or enhance the role of gas inflows and outflows) in setting galactic metallicities.
A more complete understanding of the FMR in observations is therefore critical in placing strong constraints on future simulation models' feedback implementations.

While the stellar feedback mechanisms of the four simulations analysed here are all relatively similar, each simulation implements quite different AGN feedback prescriptions (see Table~1 of \citeauthor{Wright_2024}~\citeyear{Wright_2024} and references therein).
A number of studies \citep{DeRossi_2017,Torrey_2019,van_Loon_2021,Yang_2024} show that AGN feedback can have a significant impact on the overall metallicity of galaxies.
\cite{Li_2024} use MaNGA galaxies ($0.01<z<0.15$; \citeauthor{Blanton_2017} \citeyear{Blanton_2017}) to show that the FMR itself is relatively unimpacted, compared to the MZR, by presence of AGN in galaxies.
However, the correlation between metallicity and SFR is much weaker than in galaxies without AGN.
The role of AGN in setting the FMR, and its evolution, may therefore be minimal, but it is clear that these populations deviate from the standard picture.

Mergers, on the other hand, qualitatively follow the FMR -- increased SFR corresponding to decreased metallicity -- but they are quantitatively offset from the FMR \citep{Bustamante_2020,Horstman_2021}.
Galaxy-galaxy interactions thus have a significant impact on the evolution of the baryon cycle for galaxies.
The role interacting systems play in setting the overall FMR should be set by the fraction of galaxies merging.
Generally speaking, the rate of interactions in systems increases with increasing redshift \citep[see, e.g.,][]{Lotz_2011,Rodriguez_Gomez_2015,OLeary_2021} and also changes as a function of environment \citep{L'Huiller_2012}.
Cluster galaxies tend to have higher metallicities than isolated galaxies \cite[see, e.g.,][]{Gupta_2018,Nelson_2019b,Wang_2023,Rowntree_2024}.
Moreover, both the SFR \citep{Gavazzi_2002,Poggianti_2008,Gallazzi_2021} and gas content \citep{Chung_2009,Catinella_2013} of galaxies depend on the local environment.
All of these effects can have a significant impact on the baryon cycle which may manifest as evolution of the FMR.

\section{Conclusions}
\label{sec:conclusions}

In this work, we analyse star-forming, central galaxies from the cosmological simulations Illustris, IllustrisTNG, EAGLE, and SIMBA.
We investigate the extent to which the \cite{Mannucci_2010} parameterisation of the fundamental metallicity relation (FMR) can predict the overall changes in the normalisation of the mass metallicity relation (MZR).
Furthermore, we investigate the role that SFR plays in setting the overall normalisation of the MZR (and the evolution thereof).

Our conclusions are as follows:

\begin{enumerate}[itemindent=!,labelwidth=!,labelindent=0pt, leftmargin=1.5em]
    \item We provide a new framework for understanding how the FMR predicts metallicities at high redshift: the combination of the $z=0$ fit FMR and the evolution of the SFMS combine to make predictions for the high redshift MZR (see Figure~\ref{fig:cartoon}, Section~\ref{subsec:FMR_definitions}).
    
    \item We fit a linear regression to the FMR at $z=0$ (see Equation~\ref{eqn:FMR_regressions} and also \citetalias{Garcia_2024b}).
    We find different slopes, intercepts, and $\alpha_{\rm min}$ values for each of Illustris, TNG, EAGLE, and SIMBA (Equation~\ref{eqn:FMR_regressions}).
    Additionally, we present the evolution of the SFMS in each simulation (Figure~\ref{fig:SFMS_Comp}).
    We find that the SFMS is broadly consistent between the four different simulations.
    Finally, as a point of reference, we present the true redshift evolution of the normalisation of the MZR as a function of redshift in each simulation (Figure~\ref{fig:MZR_Comp}).
    In stark contrast to the SFMS, we find that the MZR (and its evolution) is highly divergent between the four simulation models.
    
    \item By combining the $z=0$ fit FMR and evolution of the SFMS we make predictions for the redshift evolution in the MZR in each simulation (central column of Figure~\ref{fig:MZR_predictions}).
    We find that at all $z>0$ there are systematic trends with either mass or redshift (or both) in each simulation when comparing the predicted MZRs versus the true MZRs (right column of Figure~\ref{fig:MZR_predictions}).
    
    \item We define $\varepsilon_{\rm evo}$, a parameter relating the importance of SFR in setting the normalisation evolution of the MZR.
    We find that by varying $\varepsilon_{\rm evo}$ with redshift, we more closely reproduces the redshift evolution of the normalisation of the MZR (Figure~\ref{fig:varied_epsilon_predictions}).
    This result suggests that the role SFR plays in setting the normalisation may change with redshift.
    
    \item We define the static and dynamic FMRs (analogous to the ``strong'' and ``weak'' FMRs for scatter from \citetalias{Garcia_2024b}).
    The ``static'' FMR is where $\varepsilon_{\rm evo}$ is fixed with redshift meaning the the importance of SFR in setting the normalisation of the MZR is fixed with time.
    Conversely, the ``dynamic'' FMR implies that the role of SFR varies with time, indicated by $\varepsilon_{\rm evo}$ having redshift evolution.
    We perform \secondedit{a Mann-Kendall trend test}
    on the $\varepsilon_{\rm evo}$ evolution in each simulation and find that \secondedit{Illustris,} TNG and EAGLE have ``dynamic'' FMRs whereas SIMBA has a ``static'' FMR (see Section~\ref{subsubsec:static_dynamic}, Figure~\ref{fig:alpha_evo}).
    \secondedit{Using the Mann-Kendall test, in combination with Figure~\ref{fig:alpha_evo}, we further refine the dynamic FMRs of TNG and EAGLE as ``increasing'' and that of Illustris as ``decreasing''.}
    
    \item We find significant offsets from $z=0$-calibrated FMR at high redshift in TNG and EAGLE (Figure~\ref{fig:offsets}).
    In Illustris, we find no significant offsets, while in SIMBA we find that there are offsets at intermediate redshift ($z\sim3$), but not at high redshift.
    We posit that the recent \JWST{} observations showing offsets from the FMR at high-redshift (\citeauthor{Curti_2023} \citeyear{Curti_2023}; \citeauthor{Langeroodi_2023} \citeyear{Langeroodi_2023}, \citeauthor{Nakajima_2023} \citeyear{Nakajima_2023}) could possibly signal that the observed FMR is ``\secondedit{increasing} dynamic'' like the TNG and EAGLE FMRs.
    Physically, this may suggest that the evolution in galactic SFRs alone may not be enough to describe metallicity across cosmic time.
\end{enumerate}

This paper, in combination with \citetalias{Garcia_2024b}, provides a theoretical framework for a complete examination of the (potential) evolution in the FMR (see summary in Figure~\ref{fig:summary}).
While observational challenges exist in applying this framework, the understanding of whether the FMR is strong/weak and static/dynamic will offer strong constraints on future galaxy evolutionary models.
Furthermore, the broad agreement between the SFMS contrasted with the wide diversity in the MZR opens a rich parameter space for understanding what physics drives the assembly of galactic metal content.

\section*{Acknowledgements}

\finaledit{We thank the anonymous referee for their helpful comments which improved the quality of this manuscript.}
AMG acknowledges a helpful conversation about the SIMBA physical model in regards to the high redshift MZR with Desika Narayanan and Romeel Dav{\'e}.
AMG and PT acknowledge support from NSF-AST 2346977.
KG is supported by the Australian Research Council through the Discovery Early Career Researcher Award (DECRA) Fellowship (project number DE220100766) funded by the Australian Government. 
KG is supported by the Australian Research Council Centre of Excellence for All Sky Astrophysics in 3 Dimensions (ASTRO~3D), through project number CE170100013. 
\edit{RJW acknowledges support from the European Research Council via ERC Consolidator Grant KETJU (no. 818930).}

We acknowledge the Virgo Consortium for making their simulation data available. The EAGLE simulations were performed using the DiRAC-2 facility at Durham, managed by the ICC, and the PRACE facility Curie based in France at TGCC, CEA, Bruy\`eresle-Ch\^atel.

\section*{Data Availability}

All data products and analysis scripts used to support the findings in this paper are available publicly at \href{https://github.com/AlexGarcia623/does_the_fmr_evolve_simulations_2/tree/main}{https://github.com/AlexGarcia623/does\_the\_fmr\_evolve\_simulations\\\_2/tree/main}.
The raw galaxy catalogs for Illustris, TNG, EAGLE, and SIMBA are also available for public download at \href{https://www.illustris-project.org/}{https://www.illustris-project.org/}, 
\href{https://www.tng-project.org/}{https://www.tng-project.org/}, \href{https://icc.dur.ac.uk/Eagle/}{https://icc.dur.ac.uk/Eagle/}, and \href{http://simba.roe.ac.uk/}{http://simba.roe.ac.uk/}, respectively.



\bibliographystyle{mnras}
\bibliography{paper}



\appendix

\section{Dependence on specific star formation main sequence}
\label{appendix:sSFMS}

\begin{figure}
    \centering
    \includegraphics[width=\linewidth]{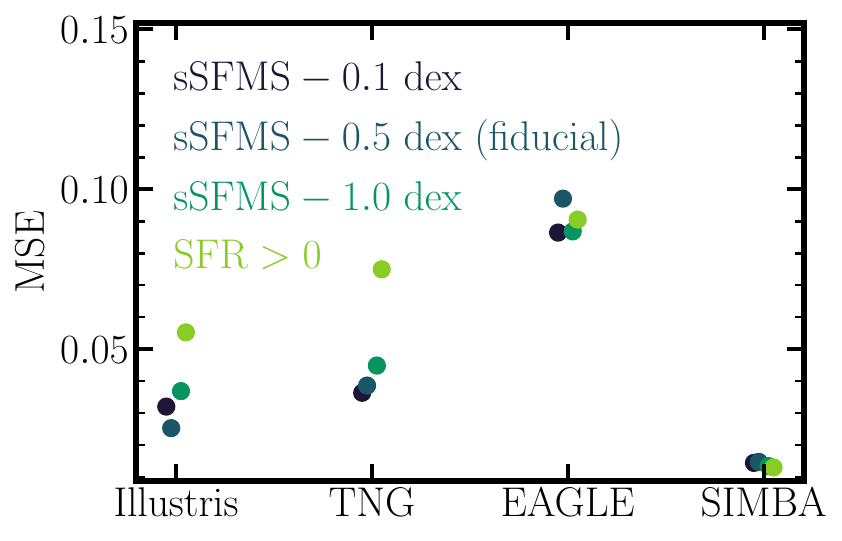}
    \caption{{\bf (Lack of) Dependence on specific Star Formation Main Sequence Cuts}. The mean-squared error ($\xi$) of the predicted MZRs for Illustris, TNG, and EAGLE using four different sSFMS cuts. From dark to light, the cuts are 0.1 dex below the median relation, 0.5 dex below the median relation, 1.0 dex below the median relation, and any ${\rm SFR} > 0$.
    }
    \label{fig:sSFMS_plot}
\end{figure}

Our galaxy selection criteria is dependent on a defined specific star formation rate main sequence (sSFMS).
We refer the reader to Section~\ref{subsec:galaxy_selection} for the complete definition; however, important for the discussion here, we omit galaxies that are 0.5 dex below the median relation.
We test the dependence of our core result on this specific definition in this appendix.

Following from \cite{Garcia_2024a}, we define three modifications to this definition of the sSFMS.
The first is omitting galaxies 0.1 dex below the median relation (more restrictive), the second is omitting galaxies 1.0 dex below the relation (less restrictive), and, finally, the third is including any galaxy that has any star formation whatsoever (least restrictive).
Figure~\ref{fig:sSFMS_plot} shows the summary metric of merit for the comparisons between the predicted MZR evolution and true MZR evolution (see Section~\ref{subsec:predicted_MZR} and right column of Figure~\ref{fig:MZR_predictions}).
We find that $\xi$ is \secondedit{very similar} between each of the variations of the threshold for selecting star-forming galaxies \secondedit{in particular for EAGLE and SIMBA}.
\secondedit{The loosest criteria (any galaxy with ${\rm SFR}>0$) in Illustris and TNG seems to increase the total derived MSE more than other specific star formation main sequence cuts.
This may suggest that systems that lie greater than $1$ dex below the main sequence but still form stars are an interesting population that does not obey the typical ``FMR''.
Regardless, owing the consistency we obtain between the different specific star formation main sequence cuts in Illustris and TNG, as well as the good agreement between EAGLE and SIMBA, we determine that our results are mostly insensitive to the detailed star formation criteria.
We do note, however, that these low star forming systems in Illustris and TNG are worthy of their own follow-up investigation.
}

\section{Oxygen and Hydrogen Mass Fractions}
\label{appendix:oxygen_abundances}

\begin{figure*}
    \centering
    \includegraphics[width=\linewidth]{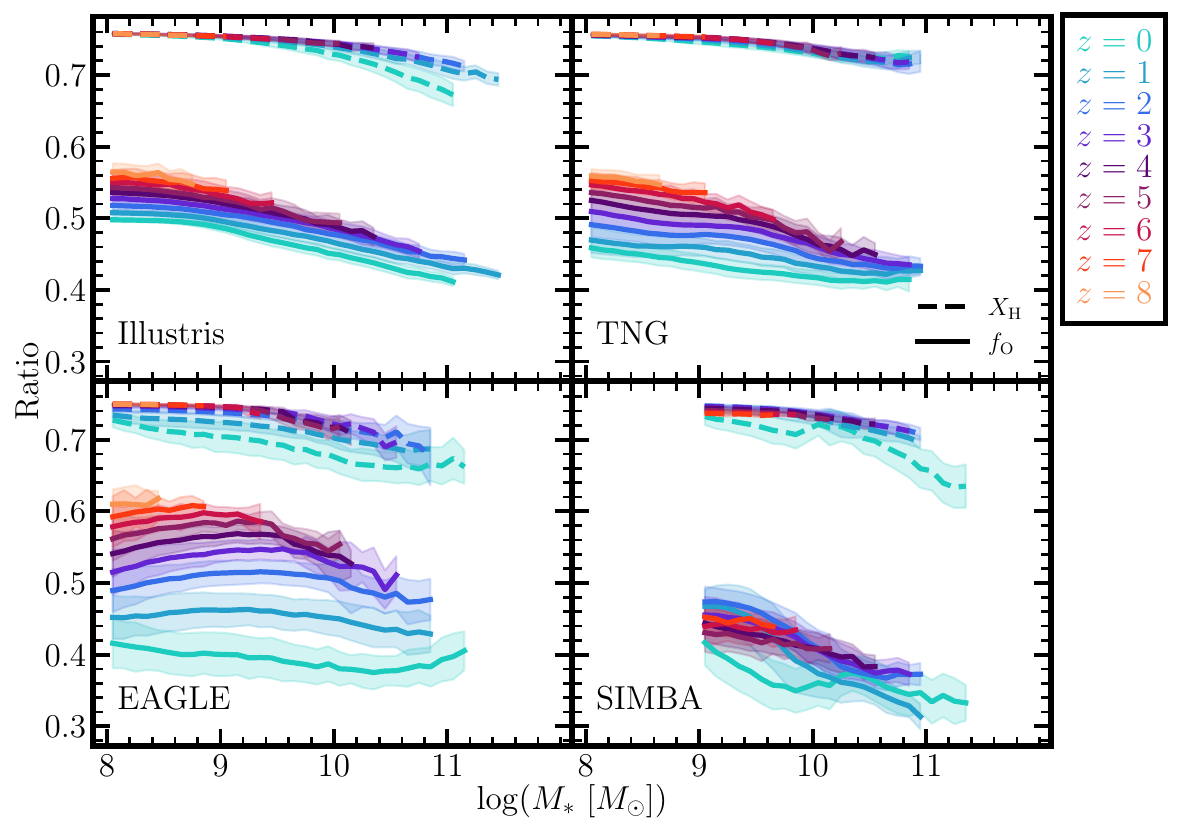}
    \caption{\secondedit{
    {\bf Evolution of mass fractions of Oxygen and Hydrogen.}
    The hydrogen to total gas mass fraction $X_{\rm H}$ (Equation~\ref{eqn:hydrogen_frac}; dashed lines) and oxygen to total metal mass fraction (Equation~\ref{eqn:oxygen_frac}; solid lines) for Illustris (top left), TNG (top right), EAGLE (bottom left), and SIMBA (bottom right).
    The shaded regions are $1\sigma$ deviations from the median relation.
    We caution that $X_{\rm H}$ and $f_{\rm O}$ are not directly additive: $f_{\rm O}$ compares only to the {\it other metals} in the system, whereas $X_{\rm H}$ compares to the total mass.
    We note that our analysis picks values of 0.5 and 0.76 for $f_{\rm O}$ and $X_{\rm H}$, respectively.
    Clearly, from this plot, these fractional values are not constant with time, mass, nor from simulation-to-simulation; however, we find that making this assumption does not significantly alter our main conclusions (see text and Figure~\ref{fig:different_Z}).
    }}
    \label{fig:oxygen_abundances}
\end{figure*}

\begin{figure*}
    \centering
    \includegraphics[width=\linewidth]{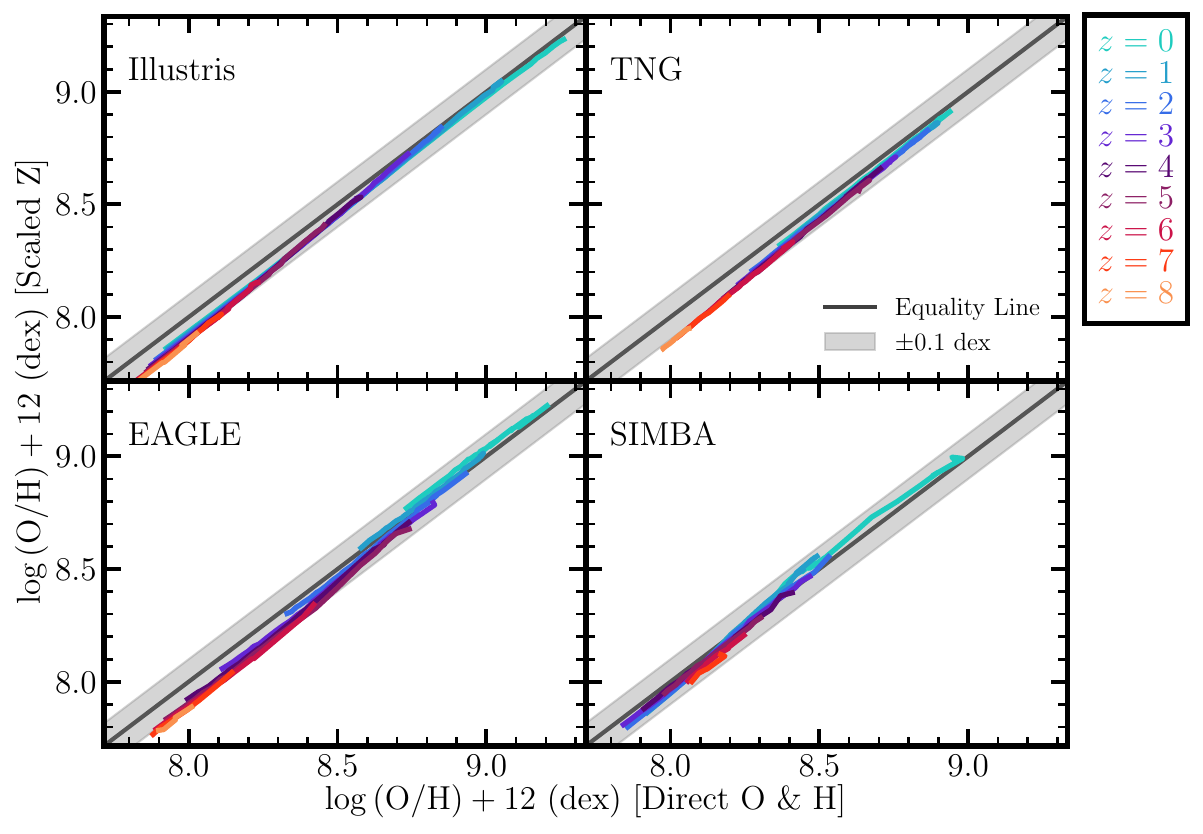}
    \caption{\secondedit{
    {\bf Comparison of $\log({\rm O/H}) + 12$ values from a scaled total metallicity versus directly tracking the oxygen and hydrogen abundances.}
    The metallicity values for Illustris (top left), TNG (top right), EAGLE (bottom left), and SIMBA (bottom right) using the mass-weighted metallicity in star forming gas via scaling the total metallicity (ordinate) versus explicitly tracing the Oxygen and Hydrogen abundances (abscissa).
    Each of the coloured lines represents data taken from the median MZR (similar to Figure~\ref{fig:MZR_Comp}) at each redshift $z=0-8$.
    The black solid line represents the one-to-one line and the shaded region represents a $\pm0.1$ dex difference from this one-to-one line (similar to typical uncertainty in observational metallicity measurements; \protect\citeauthor{Kewley_Ellison_2008} \protect\citeyear{Kewley_Ellison_2008}).
    }}
    \label{fig:different_Z}
\end{figure*}

\secondedit{
The mass fractions of Oxygen and Hydrogen are not constant with time and depend on (i) the specific enrichment pathways that a system takes and (ii) the chemical yield models from the different simulations.
To mitigate the impact of the the latter of these two effects on the derived metallicity from the simulations we choose to take $f_{\rm O}$, the oxygen-to-metal mass fraction, and $X_{\rm H}$, the total mass fraction of Hydrogen to be constants (50\% and 76\%, respectively) throughout this work (see Section~\ref{subsubsec:derived_metallicities}).
In this appendix, we investigate the impact of this decision on our conclusions.
}

\secondedit{
We begin by making specific definitions for the quantities of interest.
We define $f_{\rm O}$ as the the total mass in Oxygen in every gas cell divided by the total mass in all other metals in every cell, scaled by the metals' molecular weights.
Put more quantitatively,
\begin{equation}
    \label{eqn:oxygen_frac}
    f_{\rm O} = \cfrac{\sum_i A_{{\rm O},\, i}\, M_{{\rm gas},\, i}\, \mu_{\rm O}}{\sum_{\rm metal}\left(\sum_i A_{{\rm metal},\, i}\, M_{{\rm gas},\, i}\, \mu_{\rm metal} \right)}~,
\end{equation}
where $A_i$ is the fractional abundance of each species in cell $i$ relative to the total mass (i.e., ``GFM Metals'' from Illustris/TNG {\sc subfind} catalogs), $M_{{\rm gas},i}$ is the total gas mass in the cell, and $\mu$ is the molecular weight of the species.
We define $X_{\rm H}$ as the total mass of Hydrogen in every gas cell divided by the total gas mass
\begin{equation}
    \label{eqn:hydrogen_frac}
    X_{\rm H} = \frac{\sum_i A_{{\rm H},\,i}\,M_{{\rm gas}\,,i}}{\sum_i M_{{\rm gas},\,i}}~,
\end{equation}
where $A_{{\rm H},\,i}$ is the abundance of Hydrogen in each cell relative to the total mass of the cell.
}

\secondedit{
Figure~\ref{fig:oxygen_abundances} shows both $f_{\rm O}$ and $X_{\rm H}$ in each simulation as a function of mass from $z=0-8$.
The dashed lines represent $X_{\rm H}$ while the solid lines represent $f_{\rm O}$.
Perhaps unsurprisingly, based on the different yield implementations in each simulation, there is a significant difference between the $f_{\rm O}$ and $X_{\rm H}$ evolution between each simulation.
Illustris and TNG are the most similar (again, perhaps unsurprisingly).
In both these simulations, $f_{\rm O}$ and $X_{\rm H}$ has a mass dependence at each redshift, with more massive galaxies having lower $f_{\rm O}$ and $X_{\rm H}$ values than their low mass counterparts (though this trend is most prominent at low-$z$ in Illustris).
Moreover, at increasing redshift, both Illustris and TNG have increasing $f_{\rm O}$ with high-$z$ systems having $f_{\rm O}$ values $\sim0.08$ and $\sim0.10$ higher in Illustris and TNG (respectively).
There is virtually no trend with redshift in $X_{\rm H}$ in either simulation.
The lone exception is galaxies with $\log( M_*~[M_\odot]) > 9.0$ having lower $X_{\rm H}$ values at $z=0$.
EAGLE exhibits very little mass dependence on $f_{\rm O}$ at $z\lesssim2$, but at $z\gtrsim3$ there seems to be weak \finaledit{positive} trend with mass (although this trend also appears to invert at the larger masses in this redshift range).
Regardless, there is a clear trend of increasing $f_{\rm O}$ further back in cosmic time, ranging from $\sim0.4$ at $z=0$ to as high as $\sim0.6$ at $z=7-8$.
The $X_{\rm H}$, on the other hand, seems to have only a weak mass dependence at $\log(M_*~[M_\odot]) > 9.25$ and no redshift dependence at $z\geq2$.
At $z=0$ and $z=1$, however, the $X_{\rm H}$ fraction decreases slightly.
SIMBA has qualitatively similar behaviour to Illustris and TNG for $X_{\rm H}$, a negative trend with mass and very little evolution with time (albeit with the same exception as Illustris at $z=0$).
The overall \finaledit{mass} trend with $f_{\rm O}$ is also fairly similar in SIMBA: decreasing $f_{\rm O}$ with increasing stellar mass.
The redshift dependence on $f_{\rm O}$ is distinct, however.
The oxygen-to-metal fraction increases until $z\sim4$ and then begins to decrease at $z>4$, not unlike to the overall metallicity (see, e.g., Figure~\ref{fig:MZR_Comp}).
}

\secondedit{
Irrespective of the details of {\it how} $f_{\rm O}$ and $X_{\rm H}$ evolve with mass and redshift in the simulations, it is clear that they are not constant with either in most cases.
To quantify how significant of an effect this evolution is, Figure~\ref{fig:different_Z} shows the median metallicity derived by scaling the total metallicity by constants (using the fiducial values of $f_{\rm O}$ and $X_{\rm H}$, as in the main body of this work) as a function of the directly tracked oxygen and hydrogen abundances.
Also included in the figure is a gray band that shows $\pm0.1$ dex agreement between the two different metallicity measurements (i.e., typical uncertainty in observational metallicity measurements; \citeauthor{Kewley_Ellison_2008} \citeyear{Kewley_Ellison_2008}).
The different metallicity determinations yield remarkably similar values with the medians virtually all falling within $0.1$ dex agreement.
The level of agreement is somewhat by construction of our selection of $f_{\rm O} = 50\%$.
The oxygen-to-metal mass fraction ranges from as low as $\sim0.3$ (in SIMBA) to as high as $\sim0.625$ (in EAGLE) for some systems.
The correction required to accommodate these systems, however, is relatively modest: assuming that the deviation was only in our value of $f_{\rm O}$, these systems would only require an additive correction of $\log(0.625/0.5)\approx0.10$ dex or $\log(0.3/0.5)\approx0.22$ dex.
In most cases, though, these systems with lower (higher) $f_{\rm O}$ values tend to have higher (lower) $X_{\rm H}$ values.
As such, these corrections are usually smaller and {are typically $\lesssim$ 0.1 dex} (i.e., comparable to uncertainties in observational metallicities).
Regardless, such corrections are still not sufficient to explain all of the deviations from the true MZR and FMR predictions (which range from $0.2-0.6$ dex; see Figure~\ref{fig:MZR_predictions}).
Finally, we also briefly, mention that the scatter about the MZR is only impacted by $4-5\%$ (decreased by $\sim0.004$ dex on average) by using the directly tracked Oxygen and Hydrogen abundances.
}

\section{Different functional forms of the FMR}
\label{appendix:different_FMRs}

In the entirety of this work, we assume that the FMR takes a linear form.
The choice of linear regression is motivated primarily by recent observational works' methodology \citep{Heintz_2022,Langeroodi_2022,Nakajima_2023,Castellano_2024,Curti_2023}.
There have been a number of different FMR fits proposed in the literature, however.
In this appendix, we discuss alternative functional forms.

The summary of the findings below is that more sophisticated functional forms of the FMR better reproduce the overall shape of the MZR, yet cannot successfully reproduce the evolution of the normalisation.

\subsection{Fourth-Order Polynomial}

\begin{figure*}
    \centering
    \includegraphics[width=\linewidth]{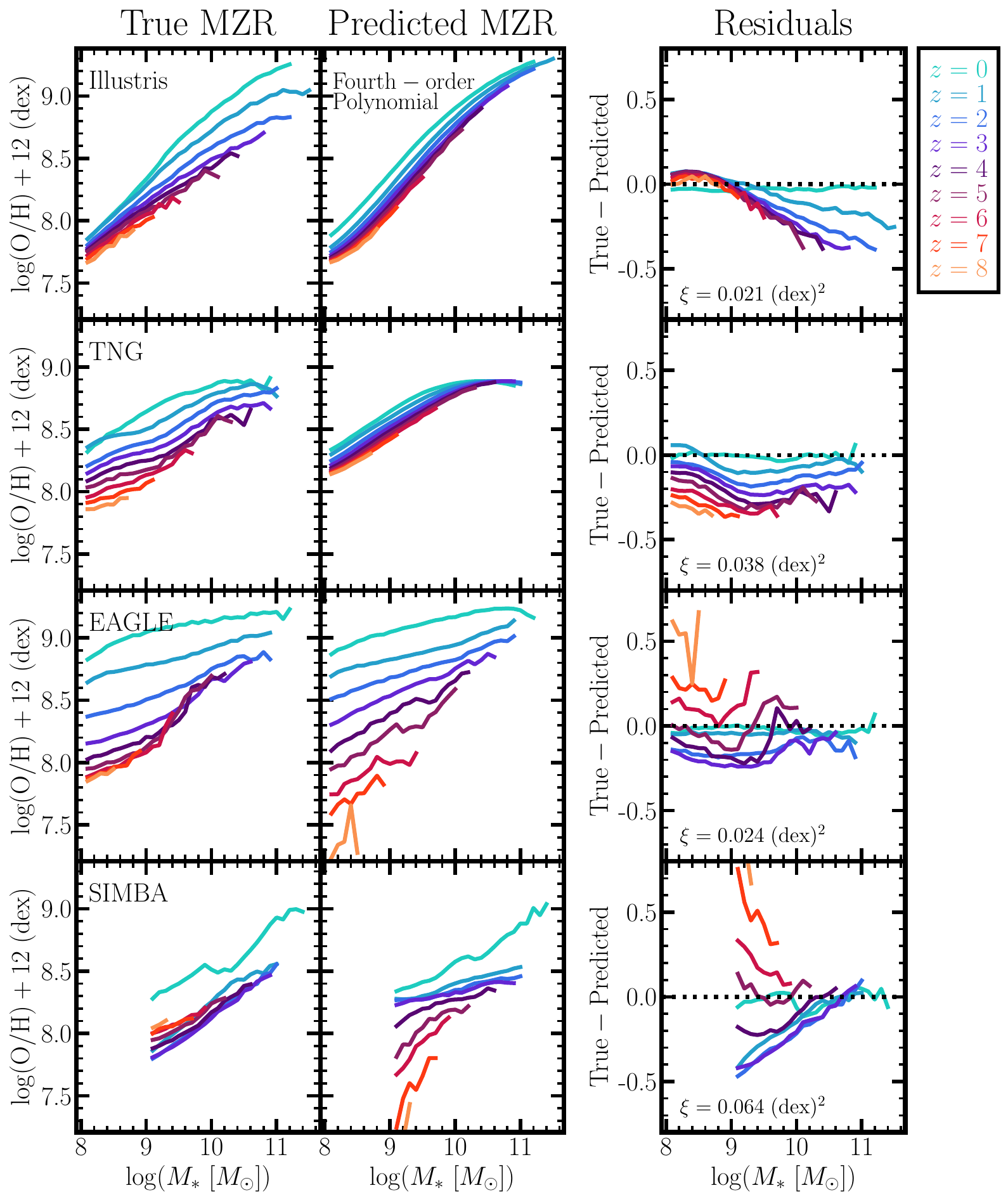}
    \caption{{\bf Same as Figure~\ref{fig:MZR_predictions} for a fourth-order polynomial $z=0$ FMR (Equation~\ref{eqn:fourth_order}).}
    Although we use a linear regression to the FMR in the main text, this choice is not uniform in the literature.
    In fact, \protect\cite{Mannucci_2010} originally proposed using a fourth-order polynomial.
    To that end, we present predictions from a fourth-order regression (see free parameters of these fits in Equations~\ref{eqn:Illustris_4}-\ref{eqn:SIMBA_4}) in the central column.
    }
    \label{fig:fourth_order_predictions}
\end{figure*}

Originally, \cite{Mannucci_2010} fit the FMR with a fourth-order polynomial, instead of a linear regression such that
\begin{equation}
    \label{eqn:fourth_order}
    Z = a y^4 + b y^3 + c y^2 + d y + e~,
\end{equation}
where $a, b, c, d,~{\rm and}~e$ are free parameters of the fit and $y=\mu_\alpha-10$\ignorespaces
\footnote{
The abscissa is re-parameterised into units of $y=\mu_\alpha-10$ to increase the numerical stability of the fitting.
}.
Those authors find that this non-linear fitting is only important at large $\mu_\alpha$ values (specifically $\mu_{0.32} < 10.5$ in their fitting) and that conclude that a piece-wise linear fit is as effective as a fourth-order polynomial in practice.
Regardless, a higher-order polynomial is used on occasion \citep[see, e.g.,][]{Bustamante_2020,Sanders_2021}.
To that end, we re-present the results for FMR predictions of the MZR (Section~\ref{subsec:predicted_MZR}) using a fourth-order polynomial here.

We determine the fourth-order FMR regressions in the same manner as the linear FMR: a linear least-squares regression to the minimum scatter $\mu_\alpha$-metallicity distribution.
We note that the $\alpha_{\rm min}$ values for the fourth-order regressions are not identical to those of the linear regressions (e.g., 0.23 for linear and 0.24 for fourth-order in Illustris).
This is since $\alpha_{\rm min}$ is determined using a fourth-order regression as well.
We show in Appendix B of \citetalias{Garcia_2024b} that this does not significantly impact the $\alpha_{\rm min}$ determination, however.
The best fit regressions are
\secondedit{
\begin{equation}
\label{eqn:Illustris_4}
\begin{alignedat}{2}
    &[\log ({\rm O/H}) + 12]_{\rm Illustris} &= 8.907+0.572y+0.201y^2\\
    & &-0.028y^3+0.029y^4~,
\end{alignedat}
\end{equation}
with $y=\mu_{0.24}-10$ for Illustris,
\begin{equation}
\label{eqn:TNG_4}
\begin{alignedat}{2}
    &\log ({\rm O/H}) + 12]_{\rm TNG} &= 8.868+0.125y-0.224y^2\\
    & &+0.010y^3+0.029y^4~,
\end{alignedat}
\end{equation}
with $y=\mu_{0.27}-10$ for TNG,
\begin{equation}
\label{eqn:EAGLE_4}
\begin{alignedat}{2}
    &[\log ({\rm rm O/H}) + 12]_{\rm EAGLE} &= 9.107+0.401y-0.089y^2\\
    & &-0.262y^3-0.240y^4~,
\end{alignedat}
\end{equation}
with $y=\mu_{0.71}-10$ for EAGLE, and
\begin{equation}
\label{eqn:SIMBA_4}
\begin{alignedat}{2}
    & [\log ({\rm O/H}) + 12]_{\rm SIMBA} &= 8.522+0.254y+0.167y^2\\
    & &+0.298y^3-0.197y^4~,
\end{alignedat}
\end{equation}
with $y=\mu_{0.66}-10$ for SIMBA.
}

Following the same procedure as outlined in Section~\ref{subsec:FMR_definitions}, we make predictions for the normalisation of the MZR now using the fourth order FMR.
The results of these predictions are shown in the central column of Figure~\ref{fig:fourth_order_predictions}.
Notably, the shape of the predicted MZR at $z=0$ is much closer to the shape of the true MZR when using a fourth-order polynomial.
Whereas the linear FMR predicted a mostly linear MZR in Figure~\ref{fig:MZR_predictions}, we see that the fourth-order polynomial has much more flexibility associated with it.
In spite of this increased flexibility, we still find that the $z=0$ calibrated FMR cannot reproduce the evolution in the normalisation of the MZR (see right column of Figure~\ref{fig:fourth_order_predictions}).
We find that $\xi$ -- the mean-squared error on the predictions -- is 0.021 (dex)$^2$ in Illustris, 0.038 (dex)$^2$ in TNG, 0.024 (dex)$^2$ in EAGLE, and \secondedit{0.064} (dex)$^2$ in SIMBA.
Comparing to the $\xi$ values of the linear FMR, we find minor improvement in Illustris, marked improvement in EAGLE, virtually no difference in TNG, \secondedit{and a significant decrease in performance in SIMBA}.
The minor improvement in Illustris is likely due to the shape of the MZR.
The true MZR in Illustris is highly non-linear at all redshifts.
We attribute the significant improvement in EAGLE to the increased level of redshift evolution predicted by the fourth-order polynomial.
Whereas the linear FMR underpredicts the redshift evolution, the fourth-order polynomial actually overpredicts the redshift evolution.
Furthermore, with the exception of EAGLE, all of the qualitative trends in the residuals for the linear FMR predictions still hold in the \secondedit{residuals} for the fourth-order predictions.
The difference in EAGLE is the aforementioned overprediction of normalisation evolution at high-redshift.

In summary, while the shape of the MZR is improved, the fourth-order polynomial FMR also cannot accurately predict the redshift evolution of the normalisation of the MZR.

\subsection{Langan et al. (2023)}

Qualitatively similar to Equation~\ref{eqn:linear} there is also the \cite{Langan_2023} form of the FMR:
\begin{equation}
    \label{eqn:Langan}
    Z(M_*, {\rm SFR}) = a \log\left(\frac{M_*}{M_{0}}\right) + b \log\left(\frac{\rm SFR}{{\rm SFR}_0}\right) + c~,
\end{equation}
where $M_0$ and ${\rm SFR}_0$ are the median stellar mass and SFR in the sample.
Although Equation~\ref{eqn:Langan} appears different than Equation~\ref{eqn:linear}, they are, in fact, equivalent.
By rearranging the constant terms, we can obtain the form of Equation~\ref{eqn:linear} with 
\begin{equation}
\begin{split}
    &m_{\rm M10} = a_{\rm L23} ~,\\
    &b_{\rm M10} = c_{\rm L23} - a_{\rm L23}\log M_0 - b_{\rm L23}\log {\rm SFR}_0~,~{\rm and}\\
    &\alpha_{\rm M10} = -\frac{b_{\rm L23}}{a_{\rm L23}}~
\end{split}
\end{equation}
where the subscript ${\rm M10}$ denotes the variable from Equation~\ref{eqn:linear} and the subscript ${\rm L23}$ denotes the variable from Equation~\ref{eqn:Langan}.
We therefore opt not to compare against this form as it would likely produce the same results as presented in Sections~\ref{subsec:predicted_MZR}.

\subsection{Curti et al. (2020)}
\label{appendix:Curti}

\begin{table*}
    \centering
    \begin{tabular}{lx{0.125\linewidth}x{0.125\linewidth}x{0.125\linewidth}x{0.135\linewidth}x{0.125\linewidth}}
         \toprule
          & {\bf Illustris} & {\bf TNG} & {\bf EAGLE}  & {\bf SIMBA} & {\bf \citetalias{Curti_2020}}\\\midrule
          $Z_0\bluedagger$ & \secondedit{9.36}  \secondedit{$\pm$ 0.04} & \secondedit{8.88} \secondedit{$\pm$ 0.01} & \secondedit{9.23} \secondedit{$\pm$ 0.02} & \secondedit{292.32} \secondedit{$\pm$ 32407.5} & 8.779 \secondedit{$\pm$ 0.005} \\
          $\gamma$         & 1.10  \secondedit{$\pm$ 0.12} & 0.24 \secondedit{$\pm$ 0.17} & 1.13 \secondedit{$\pm$ 0.79} & \secondedit{0.51}  \secondedit{$\pm$ 49.67  } & {0.31}  \secondedit{$\pm$ 0.01 } \\
          $\beta$          & 1.19  \secondedit{$\pm$ 0.08} & 0.98 \secondedit{$\pm$ 0.46} & 0.83 \secondedit{$\pm$ 0.32} & \secondedit{0.00}  \secondedit{$\pm$ 0.02   } & 2.1   \secondedit{$\pm$ 0.4  } \\
          $m_0$            & 10.31 \secondedit{$\pm$ 0.05} & 9.57 \secondedit{$\pm$ 0.16} & 9.17 \secondedit{$\pm$ 0.91} & \secondedit{38.77} \secondedit{$\pm$ 155318.59} & 10.11 \secondedit{$\pm$ 0.03 } \\
          $m_1$            & 0.42  \secondedit{$\pm$ 0.05} & -0.45\secondedit{$\pm$ 0.95} & 0.58 \secondedit{$\pm$ 0.15} & \secondedit{-0.47} \secondedit{$\pm$ 0.48   } & 0.56  \secondedit{$\pm$ 0.01 }\\\bottomrule
    \end{tabular}
    \caption{{\bf Best fit parameters of Equation~\ref{eqn:Curti_FMR}.}
    The best fit parameters for the \protect\cite{Curti_2020} functional form of the FMR for Illustris, TNG, EAGLE, and SIMBA.
    The right-most column are the regression parameters from \protect\citeauthor{Curti_2020} (\protect\citeyear{Curti_2020}; their Table 6).
    \secondedit{The uncertainty on the fit parameters (of the simulations) are the square-root of the diagonal elements of the covariance matrix.}
    $\bluedagger$Following from \protect\cite{Curti_2020}, $Z_0$ is taken as a constant determined from the MZR fitting (i.e., from Equation~\ref{eqn:Curti_MZR}), see Section~\ref{appendix:Curti} for more details.
    }
    \label{tab:Curti_table}
\end{table*}

\begin{figure*}
    \centering
    \includegraphics[width=\linewidth]{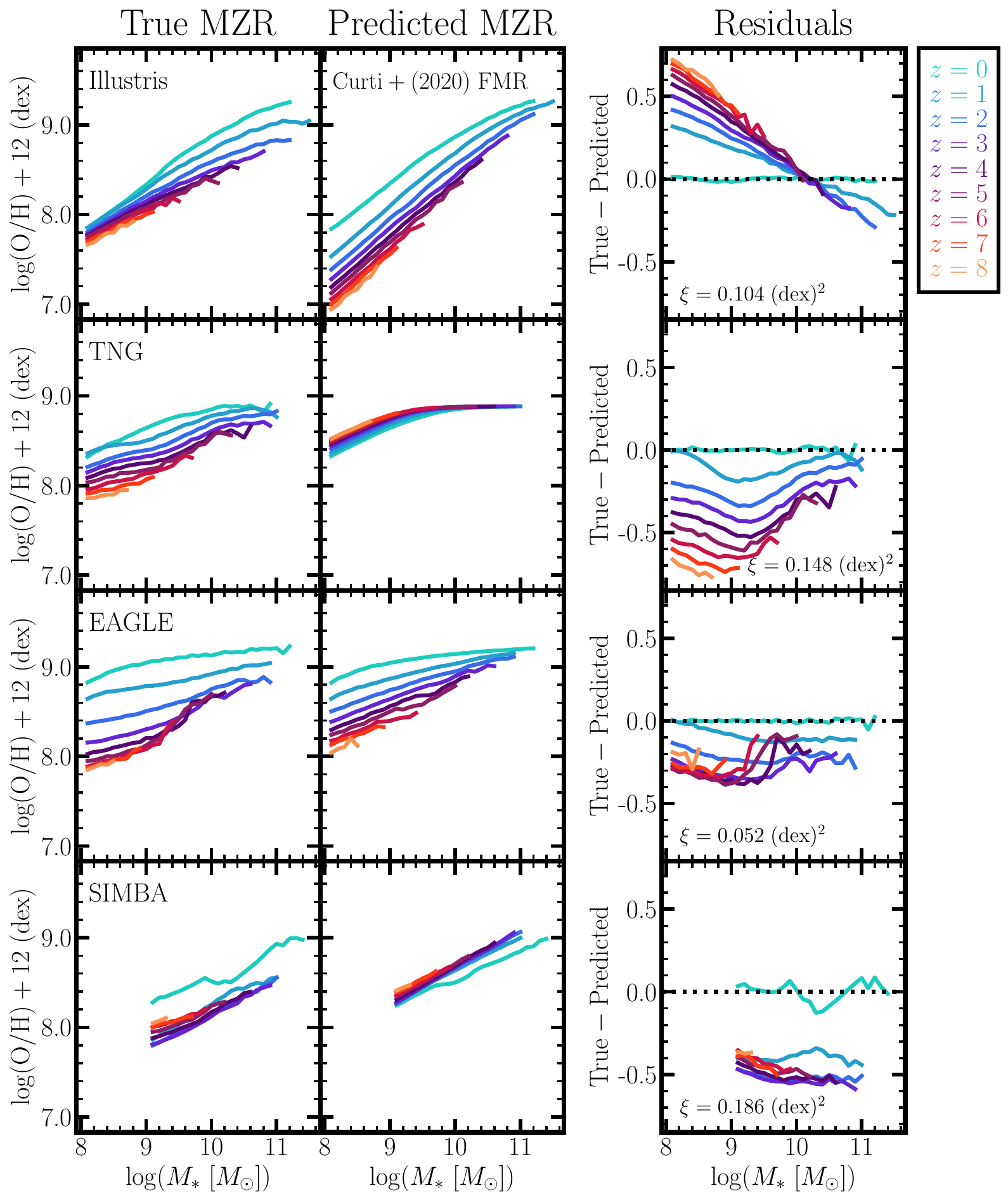}
    \caption{{\bf Same as Figure~\ref{fig:MZR_predictions} but for the \protect\cite{Curti_2020} FMR (Equation~\ref{eqn:Curti_FMR}).} 
    We caution that the $z=0$ MZRs in TNG and SIMBA are not well-conditioned to this functional form.
    Therefore the predictions made here in TNG and SIMBA should be taken lightly.
    }
    \label{fig:MZR_predictions_Curti}
\end{figure*}

\cite{Curti_2020} introduce a new functional form of the FMR that stems from a functional form of the MZR with the addition of a SFR dependence.
The functional form of the MZR, as introduced by \cite{Curti_2020}, is of the form
\begin{equation}
    \label{eqn:Curti_MZR}
    Z(M_*) = Z_0 - \frac{\gamma}{\beta}\log\left(1 + \left(\frac{M_*}{M_0}\right)^{-\beta}\right)~,
\end{equation}
where $Z_0$ is the value the MZR asymptotically approaches at high stellar mass, $M_0$ is the turnover mass, $\gamma$ is the low mass power-law index, and $\beta$ controls the rate at which the transition from power-law to asymptotic metallicity occurs.
The FMR of \cite{Curti_2020} is extended by recognising that the mass at which the MZR begins to flatten changes in thin SFR bins.
The SFR dependence is integrated such that
\begin{equation}
    \label{eqn:Curti_FMR}
    Z(M_*, {\rm SFR}) = Z_0 - \frac{\gamma }{ \beta } \log\left( 1 + \left( \frac{M_*}{M_0({\rm SFR})}\right)^{-\beta} \right) ~,
\end{equation}
where $M_0({\rm SFR})$ is the relationship between the turnover mass in the MZR as a function of SFR bin.
More specifically, $M_0({\rm SFR})$ is defined by a linear regression in $\log M_0$-$\log {\rm SFR}$ space: $\log M_0({\rm SFR}) = m_0 + m_1 \log {\rm SFR}$.

To fit the FMR parameters\ignorespaces
\footnote{
We note that we also fit the MZR using the same methodology to determine $Z_0$ which is held fixed in the FMR regression.
\label{ftn:Curti}
}
we use the non-linear least squares curve fit routine provided by {\sc scipy} \citep{Scipy}.
We provide initial guesses to the curve fitting routine as the best fit parameters from \citeauthor{Curti_2020} (\citeyear{Curti_2020}; see right-most column of Table~\ref{tab:Curti_table}) and set the maximum number of iterations to 5000.
We note, however, that owing to the mismatch between the shape of the $z=0$ MZR in SIMBA and the observed MZR (discussed more below), we increase the max iterations for SIMBA to 1,000,000 to ensure convergence.
The parameters of the FMR fit are presented in Table~\ref{tab:Curti_table} for Illustris, TNG, EAGLE, and SIMBA.
\secondedit{The uncertainty on our fit parameters in the table come from the square root of their variance (given by the diagonal elements of the covariance matrix).}

For the most part, we obtain qualitatively reasonable results from the fitting methodology.
For example, the asymptotic metallicity values in Illustris, TNG, and EAGLE appear consistent with the $z=0$ MZR as presented throughout this work (see, e.g., Figure~\ref{fig:MZR_Comp}).
Notable exceptions to the qualitative agreement exist, however.
We specifically note that (i) $m_1$ in TNG has the opposite sign of observations, Illustris and EAGLE \secondedit{(though there is a large uncertainty on this parameter)} and (ii) the parameters in SIMBA vary significantly from any other simulation or observations \secondedit{(with extremely large uncertainties)}.

The negative $m_1$ in TNG is caused by low SFR bins not having strong turnovers.
By fitting SFR bins of width 0.15 dex with Equation~\ref{eqn:Curti_MZR}, we find values of $M_0$ that range from $20-80$ in some of lower SFR bins.
This implies that the turn over mass of the MZR is at stellar masses of $10^{20}-10^{80} M_\odot$ in these SFR bins, which is clearly unphysical.
Rather, we interpret this as these SFR bins do not have a strong (or, perhaps, any) turnover mass.
This is qualitatively consistent with the findings of \citetalias{Garcia_2024b}, wherein we determine that the $z=0$ MZR has a much weaker correlation with SFR than both 
(i) its higher-redshift counterparts and 
(ii) Illustris and EAGLE at $z=0$.

SIMBA's $z=0$ MZR parameters
are all poorly conditioned.
The failure of the fitting methodology in SIMBA is likely due to the median MZR in SIMBA following a \secondedit{significantly different} shape than the observed MZR.
The \cite{Curti_2020} functions explicitly parameterise the key features of the MZR.
The shape of the MZR in SIMBA does not follow the proto-typical structure that the functions parameterise.
This non-standard structure is reflected in the obtained fitting parameters: the metallicity the MZR \secondedit{asymptotes} to, $Z_0$, is $\sim\!300$ and the rate at which the power-law transitions to the flattening, $\beta$, is 0.
We therefore caution that interpreting predictions from the \cite{Curti_2020} FMR is not straightforward in SIMBA.

In spite of these irregularities, we present the predictions for the evolution in the normalisation of the MZR in Figure~\ref{fig:MZR_predictions_Curti} (central column).
We find that Illustris and EAGLE predict evolution such that metallicities at higher-redshift are lower than at $z=0$.
Conversely, SIMBA and TNG predict that higher redshift galaxies should be more metal {\it rich} than lower redshift.
The predictions in SIMBA and TNG are likely driven by the features discussed in the previous two paragraphs.
We show the residuals of these predictions in the right column of Figure~\ref{fig:MZR_predictions_Curti}.
As in all previous comparisons, we use the mean-squared error ($\xi$) as a summary metric: $0.104$ (dex)$^2$ in Illustris, 0.148 (dex)$^2$ in TNG, 0.052 (dex)$^2$ in EAGLE, and \secondedit{0.186} (dex)$^2$ in SIMBA.
These predictions are significantly worse than the linear FMR predictions in all but EAGLE.
The difference in Illustris is the more significant underprediction of metallicities in the lowest mass bins at $z>0$.
The predictions for EAGLE are just slightly too metal rich at higher redshift.
Interestingly, the predicted MZRs at all $z>3$ are nearly a constant offset of $\sim\!0.25$ dex.

In summary, we find that the \cite{Curti_2020} functional form of the FMR also cannot accurately predict the redshift evolution of the normalisation of the MZR.

\section{Weak FMR in SIMBA}
\label{appendix:SIMBA_strong_weak}

\begin{figure*}
    \centering
    \includegraphics[width=\linewidth]{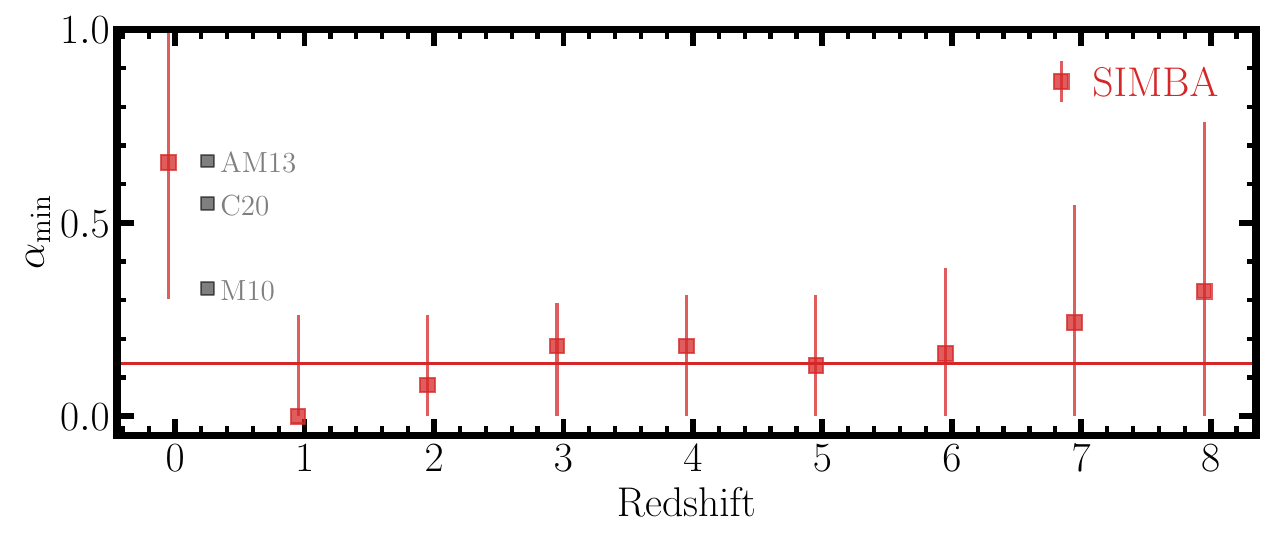}
    \caption{{\bf $\alpha_{\rm min}$ values as a function of redshift in SIMBA.}
    Following the procedure from \protect\citetalias{Garcia_2024b}, we determine $\alpha_{\rm min}$ -- a parameter that describes the importance of SFR in setting the scatter about the MZR -- at $z=0-7$ in SIMBA.
    The uncertainty on $\alpha_{\rm min}$ is determined by finding values of $\alpha$ that reduce the scatter to within 5\% of the minimum.
    The solid line is a weighted mean of all the $\alpha_{\rm min}$ values (see Section~\ref{appendix:SIMBA_strong_weak} for more details).
    We find that the average $\alpha_{\rm min}$ deviates significantly from that of the $z=0$ value.
    We therefore conclude that the FMR is ``weak'' in SIMBA.
    The gray squares are observational values of $\alpha_{\rm min}$ in the literature (\citeauthor{Mannucci_2010} \citeyear{Mannucci_2010} as M10, \citeauthor{Andrews_Martini_2013} \citeyear{Andrews_Martini_2013} as AM13, and \citeauthor{Curti_2020} \citeyear{Curti_2020} as C20).
    }
    \label{fig:alpha_min_SIMBA}
\end{figure*}

\citetalias{Garcia_2024b} did not include analysis for SIMBA.
Therefore, for the purposes of Section~\ref{subsec:taking_stock}, we briefly discuss SIMBA in the context of the scatter about the MZR.
More specifically, we examine whether or not $\alpha_{\rm min}$ -- the parameter relating the importance of scatter about the MZR -- evolves with redshift in SIMBA.

Figure~\ref{fig:alpha_min_SIMBA} shows $\alpha_{\rm min}$ at $z=0-8$ in SIMBA.
We note that we derive $\alpha_{\rm min}$ using identical methodology of \citetalias{Garcia_2024b}: vary $\alpha$ from 0 to 1 in steps of 0.01 and find which $\alpha$ value produces the least scatter in the $\mu_\alpha$-metallicity relation.
Similarly, we compute the uncertainty using a 5\% deviation from the minimum scatter.
At $z=0$, we find an $\alpha_{\rm min}$ value of \secondedit{0.66}, which is quite similar to that found in \citeauthor{Andrews_Martini_2013} (\citeyear{Andrews_Martini_2013}), though it varies from the \citeauthor{Mannucci_2010} \citeyear{Mannucci_2010} value and 1.0 (see \citeauthor{Andrews_Martini_2013} \citeyear{Andrews_Martini_2013} for a discussion of the impact of metallicity diagnostic on the derived $\alpha$ value).
At all $z>0$, we find $\alpha$ values that are significantly reduced compared to that of $z=0$.
In fact, at all other redshifts the uncertainty of $\alpha_{\rm min}$ includes $\alpha = 0.0$.
This lack of dependence on SFR was noted in the original SIMBA analysis \citep[see][their Section 3.5 and Figure 9]{Dave_2019}.
\citeauthor{Dave_2019} attribute this lack of dependence to a large population of quenched galaxies at $z\sim2.3$.
Regardless, it is interesting to note that $\alpha=0.0$ implies that there is not a significant FMR for scatter in SIMBA at high-redshift (in particular, at $z=1$).
It is unclear what might drive the FMR to ``turn-on'' and then ``turn-off'' again.
While this is certainly an interesting result, we leave further investigation of the SIMBA physical model to another work.

For the purposes of this work, we perform a one-sample $t$-test to determine whether or not $\alpha_{\rm min}$ varies as a function of redshift.
As in \citetalias{Garcia_2024b}, we compute a weighted mean of the $\alpha_{\rm min}$ values by normalising the reciprocal of the squared uncertainty.
Furthermore, we normalise the resulting $t$-statistic by the sum of the squared uncertainties.
We obtain a $t$-statistic of \secondedit{-8.989} and a $p$-value of \secondedit{$1.87\times10^{-5}$}.
Therefore, at the 0.05 confidence level, we reject the null hypothesis that $\alpha_{\rm min}$ is unevolving and conclude that the FMR for scatter in SIMBA is weak.


\bsp	
\label{lastpage}
\end{document}